\documentclass[aps,twoside,twocolumn,pra]{revtex4-2}

\usepackage{mathrsfs}

\usepackage{calligra}
\usepackage{graphicx}
\usepackage{dcolumn}
\usepackage{bm}
\usepackage{braket}
\usepackage{bbold} 
\usepackage{fancyvrb}       
\usepackage{amsfonts}  
\usepackage{mathtools} 
\usepackage[
    colorlinks=true,       
    urlcolor=blue,         
    citecolor=blue         
]{hyperref}

\usepackage{orcidlink}

\renewcommand{\d}{\ensuremath{\mathrm{d}}}

\newcommand{\M}{\mathcal{M}}
\newcommand{\R}{\mathbb{R}}

\newcommand{\Ric}{R} 

\newcommand{\iu}{\mathrm{i}} 

\newcommand{\llambda}{{\bm\lambda}}
\newcommand{\aalpha}{{\bm\alpha}}

\newcommand{\HH}{\hat{H}}

\renewcommand{\Im}{\text{Im}}
\renewcommand{\Re}{\text{Re}}

\def\scal#1#2{\langle#1|#2\rangle}
\def\matr#1#2#3{\langle#1|#2|#3\rangle}

\begin{document}

\title{Quantum geometry in many-body systems with precursors of criticality}

\author{Jan St{\v r}ele{\v c}ek\,\orcidlink{0000-0003-1752-8891}}%
\email{jan.strelecek@matfyz.cuni.cz}
\author{Pavel Cejnar\,\orcidlink{0000-0003-4206-3555}}%
\email{pavel.cejnar@matfyz.cuni.cz}

\affiliation{Institute of Particle and Nuclear Physics, Faculty of Mathematics and Physics, \\
Charles University, V Hole{\v s}ovi{\v c}k{\'a}ch 2, 180 00 Prague, Czechia}

\date{\today}

\begin{abstract}
We analyze the geometry of the ground-state manifold in parameter-dependent many-body systems with quantum phase transitions (QPTs) and describe finite-size precursors of the singular geometry emerging at the QPT boundary in the infinite-size limit.   
In particular, we elucidate the role of diabolic points in the formation of first-order QPTs, showing that these isolated geometric singularities represent seeds generating irregular behavior of geodesics in finite systems.
We also demonstrate that established approximations, namely the mean field approximation in many-body systems composed of mutually interacting bosons and the two-level approximation near a~diabolic point, are insufficient to provide a~reliable description of geometry.
The outcomes of the general analysis are tested and illustrated by a~specific bosonic model from the Lipkin-Meshkov-Glick family.
\end{abstract}


\date{\today}
\maketitle


\section{Introduction}
\label{sec:introduction}

The use of geometric perspective has been very fruitful in many branches of physics, including quantum theory \cite{Nakahara03}.
A great achievement of the geometric approach in quantum mechanics was Berry's discovery of the geometric phase acquired in the adiabatic evolution of quantum systems \cite{Berry84,Wilczek88}.
Independently, Provost, Vallee \cite{Provost80} and Wootters \cite{Wooters81} studied the concept of distance in the space of quantum states.
Their approach can be considered as a~generalization of the Fisher information metric from the space of probability distributions to the complex projective Hilbert space (\lq\lq space of rays\rq\rq), being essentially equivalent to a~much earlier definition of the Fubini-Study metric \cite{Bengttson06}.
These results were further elaborated by Ashtekar, Schilling and others, see, e.g., Refs.\,\cite{Ashtekar99,Brody01,Facchi10}, in their attempts to formulate quantum mechanics in a~fully geometric language.

Quantum geometric concepts are commonly used in adiabatically driven systems and adiabatic state preparation techniques, see for instance Refs.\,\cite{Anandan90,Miyake01,Brody03,Rezakhani09,Jones10,Tomka16,Kolodrubetz17,Bleu18,Bukov19,Amghar20,Bhandari20,Li22,Atif22,Matus23a,Matus23b,Cejnar23,Matus24,Mainersen24}.
Research along these lines is important primarily in the context of adiabatic quantum computation \cite{Albash18} and related quantum-information procedures, but it is relevant also for gate quantum algorithms \cite{Atif22,Nielsen10}.
Some of the analyses \cite{Anandan90,Rezakhani09,Tomka16,Kolodrubetz17,Bukov19,Li22,Mainersen24} propose that driving along geodesics (paths with minimal geometric length in the parameter space) is, in some sense, optimal. 
Other studies \cite{Matus23a,Matus23b,Matus24} demonstrate that although geodesics do not yield the maximal fidelity of target state preparation for coherent driving protocols, some other geometry-inspired paths represent promising candidates for optimal driving strategies.
Moreover, as shown in Ref.\,\cite{Cejnar23}, the superior role of geodesics can be re-established in open systems which decohere with a~bath or interact in a~specific way with an artificial ancilla.

This research intersects with the study of quantum phase transitions (QPTs), i.e., nonanalytic changes of the ground state at some critical values of control parameters \cite{Sachdev99,Carr11}.
Indeed, in most state-preparation tasks, the system of interest must cross a~finite-size precursor of a~QPT from uncorrelated (factorized) to highly correlated (entangled) ground states.
Quantum geometry of systems with QPT was first considered in Refs.\,\cite{Zanardi07,Venuti07}, with the emphasis on consequences for quantum information theory, and then elaborated in numerous studies of specific systems; see, e.g., Refs.\,\cite{Rezakhani10,Ma10,Dey12,Kumar12,Kolodrubetz13,Kumar14,Henriet18,Tan19,Carollo20,Jaiswal21,Liska21,Gonzales21,Gutierrez21,Gutierrez22,Sousa23,Pal23,Lambert23,Bukov14}.
Among the problems studied in these works, let us mention the description of QPT-induced singularities of the metric tensor and their precursors in finite systems and the existence of simple invariant geometries characterizing the ground-state quantum phases on both sides of a~QPT. Note that analogous geometric approaches are applied also in the field of thermodynamic phase transitions \cite{Ruppeiner95,Aman03}.
Despite the tremendous progress of all these efforts, many problems in the geometric description of quantum critical systems remain inconclusive and open to further research.

In this paper, we address several important aspects of the ground-state geometry in finite-size realizations of quantum critical systems.
It has been shown that QPT boundaries in the parameter space form impenetrable barriers for geodesics \cite{Kumar12}.
Here we investigate how this behavior, which emerges only in the infinite-size limit, is being gradually approached in finite systems.
One of the apparent suspects, which we investigate in detail, are the diabolic points (DPs), i.e., isolated degeneracies of the ground and first-excited states that sometimes appear as finite-size precursors of more robust QPT degeneracies.
As the DPs represent geometrical singularities, which (due to their ability to absorb some geodesics) can be compared to black holes, they may play a~key role in the emergence of first-order QPTs.
However, the geometric barrier associated with a~QPT can also arise without DPs, via creating a~zone with a~high geometric curvature which deviates all geodesics along (or away from) the QPT boundary.
We will present a~model in which both these mechanisms are relevant.
Other topics addressed in this paper concern the invariant geometries associated with individual ground-state phases (no simple geometries are identified in our model) and the mean field approximation of the geometric quantities (this approximation turns out fatally insufficient for geometric purposes).

We consider a~general quantum system with Hamiltonian $\hat{H}^{(N)}(\llambda)$, where $N$ is the system size parameter (typically, a~number of elementary constituents---qubits, particles etc.) and ${\llambda=(\lambda_1,\dots,\lambda_D)}$ represents a~$D$-dimensional set of control parameters that belong to a~certain domain $\mathbb{\Omega}\subseteq\R^D$ called parameter space.
We particularly focus on the cases with ${D=2}$, which is the lowest dimension for which DPs generically appear in the parameter space in the form of isolated singularities.
The ${N\to\infty}$ limit Hamiltonian $\hat{H}^{(\infty)}(\llambda)$ is supposed to exhibit emergent QPT phenomena, which for ${D=2}$ show up along 1-dimensional critical curves (quantum phase separatrices) in the parameter space.

We first present some general insight into quantum geometry, which is independent of the specific many-body system under study.
Then, to illustrate this acquired understanding, we analyze the geometry of a~simple interacting boson system, equivalent to a~fully connected system of~$N$ qubits, that shows chains of DPs in finite-$N$ realizations and first- and second-order QPTs in the infinite-$N$ limit.
The concrete plan is as follows:
In Section~\ref{sec:geoground}, we recapitulate general terms of the geometric description of the ground-state manifold and discuss from a~new perspective the geometric implications of DPs and QPTs.
In Section~\ref{sec:multi_qubit} and Appendix~\ref{sec:bosco}, we outline general bosonic systems with a~finite number of degrees of freedom, summarize the mean field approximation in such systems and its drawbacks in the description of quantum geometry, and introduce the specific model employed in the subsequent case study.
In Section~\ref{sec:geomodel}, we present a~comprehensive numerical analysis of the geometric properties of the ground-state manifold for the specific model from Sec.\,\ref{sec:multi_qubit}. 
Finally, in Section~\ref{sec:conc}, we summarize the results and present conclusions.


\section{Geometry of ground states}
\label{sec:geoground}

\subsection{Metric tensor on the ground-state manifold}

As indicated above, we consider a~system with a~Hamiltonian operator $\HH^{(N)}(\llambda)$ depending explicitly on $D$ real control parameters ${\llambda\equiv\{\lambda_{\mu}\}_{\mu=1}^D}$ and on an adequately defined size parameter $N$ (e.g., a~conserved number of particles). 
The spectrum of energies $E_k^{(N)}(\llambda)$ with $k\!=\!{0,1,\dots}$ (we require ordering ${E_0^{(N)}\leq E_1^{(N)}\leq\dots}$) is assumed to be discrete and non-degenerate, possibly except some singular parameter subdomains (zero-measure subsets of the parameter space $\mathbb{\Omega}$) where degeneracies may occur.
The eigenvectors corresponding to eigenvalues $E^{(N)}_k(\llambda)$ are denoted as $\ket{\psi^{(N)}_k(\llambda)}$. 

The ground state manifold $\M^{(N)}_0$ is defined as a~set of normalized ground states $\ket{\psi^{(N)}_0(\llambda)}$ for all $\llambda\in\mathbb{\Omega}$. 
Vectors in $\M^{(N)}_0$ can be multiplied by arbitrary parameter-dependent phase factors $e^{\iu\varphi(\llambda)}$ (this gives the manifold a~fibred structure), but the theory must be completely invariant under such gauge transformations.
Note that all considerations can be easily generalized to the manifolds $\M^{(N)}_k$ of the higher eigenstates $\ket{\psi^{(N)}_k(\llambda)}$, but below, we will be dealing only with $\M^{(N)}_0$.

The element of distance on $\M^{(N)}_0$ can be defined in terms of the Provost-Vallee metric \cite{Provost80}
\begin{equation}
    \d\ell^2=1-\left|\braket{\psi^{(N)}_0(\llambda\!+\!\d\llambda)\bigl|\psi^{(N)}_0(\llambda)}\right|^2
    = g^{(N)}_{\mu\nu}(\llambda)\,\d\lambda_{\mu}\,\d\lambda_{\nu},
    \label{eq:distanc}
\end{equation} 
where ${\d\lambda_{\mu}\equiv\d\llambda=(\d\lambda_1,\dots,\d\lambda_D)}$ is an infinitesimal change of control parameters (contrary to the usual notation, we use only lower indices in all expressions). 
The first part of this formula expresses the probability that for a~system prepared in the initial ground state at the parameter point $\llambda$, an infinitesimal jump of parameters $\llambda\to{\llambda+\d\llambda}$ induces a~transition to any of excited states at the final parameter point.
If the ground state $\ket{\psi^{(N)}_0(\llambda\!+\!\d\llambda)}$ is the same (up to a~phase factor) as $\ket{\psi^{(N)}_0(\llambda)}$ or orthogonal to it, the distance is zero or one, respectively.
The second part of Eq.\,\eqref{eq:distanc} introduces the metric tensor $\{\,g^{(N)\,}_{\mu\nu}\}\equiv\bm{g}^{(N)}$, which forms a~$D\times D$ real matrix describing increments of the distance along various directions in the parameter space.
Note that we are using here the summation convention for double Greek indices corresponding to the coordinate components.

The quantum metric tensor can be deduced from a~more general object called the geometric tensor \cite{Berry88}.
It is defined by
\begin{eqnarray}
    G^{(N)}_{\mu\nu}(\llambda)=&&\left\langle\partial_\mu\psi^{(N)}_{0}(\llambda)\left|\hat{P}^{(N)}_{0\perp}(\llambda)\right|\partial_\nu\psi^{(N)}_{0}(\llambda)\right\rangle
\label{eq:geo},\\
    &&\hat{P}^{(N)}_{0\perp}(\llambda)=\hat{I}-\ket{\psi^{(N)}_0(\llambda)}\bra{\psi^{(N)}_0(\llambda)}
\label{eq:projpen}
\end{eqnarray}
where $\hat{P}^{(N)}_{0\perp}(\llambda)$ (with $\hat{I}$ denoting the unit operator) stands for a~projector to the complement of the ground-state subspace at given $\llambda$ and $\partial_\mu X\equiv\frac{\partial}{\partial\lambda_{\mu}}X$.
The metric tensor coincides with the symmetric part of the geometric tensor, or equivalently
\begin{equation}
 g^{(N)}_{\mu\nu}(\llambda)=\Re\,G^{(N)}_{\mu\nu}(\llambda).
 \label{eq:gRe}
\end{equation}
The antisymmetric (imaginary) part of \eqref{eq:geo}, or more precisely ${F^{(N)}_{\mu\nu}=-2\,\Im\,G^{(N)}_{\mu\nu}}$, represents the Berry curvature tensor describing the geometric phases acquired by the ground state subject to adiabatic driving \cite{Berry84}.

We can evaluate the geometric tensor from the perturbation theory, yielding a~formula
\begin{equation}
    G^{(N)}_{\mu\nu}\!=\!\sum_{k>0}\frac{
    \matr{\psi^{(N)}_0}{\partial_\mu\HH^{(N)}}{\psi^{(N)}_k}\matr{\psi^{(N)}_k}{\partial_\nu\HH^{(N)}}{\psi^{(N)}_0}
    }{\bigl(E^{(N)}_k-E^{(N)}_0\bigr)^2}
    \label{eq:geopert}
\end{equation}
(here we hide dependencies of all quantities on $\llambda$), which is more suitable for numerical calculations as it replaces the derivatives of state vectors in Eq.\,\eqref{eq:geo} by computationally simpler derivatives of the Hamiltonian.

In systems composed of $N$ elementary constituents, for which the mean field method provides a~valid approximation for large $N$, the nearly factorized form of the large-$N$ scalar product in Eq.\,\eqref{eq:distanc} implies that  the metric tensor scales as 
\begin{equation}
    \bm{g}^{(N)}(\llambda)\approx N\bm{g}(\llambda),
    \label{eq:intens}
\end{equation}
where $\bm{g}(\llambda)=\lim_{N\to\infty}\bm{g}^{(N)}(\llambda)/N$ represents an infinite-size limit of the scaled quantity, which we call {\em intensive metric tensor}. 
We see that all distances generated by the metric tensor $\bm{g}^{(N)}$ diverge if the size grows infinitely large, which is a~consequence of the known fact that even infinitesimally close ground states $\ket{\psi^{(N)}_0(\llambda)}$ and $\ket{\psi^{(N)}_0(\llambda\!+\!\d\llambda)}$ of an interacting $N$-body system become orthogonal for ${N\to\infty}$ \cite{Anderson67,Zanardi06}.
This effect, sometimes called the \lq\lq orthogonality catastrophe\rq\rq, implies that geometry in the truly infinite-size limit can be studied only via the intensive metric tensor $\bm{g}$.

\subsection{Geodesics, scalar curvature and geometry}
\label{sec:geo}

Having defined the metric tensor on the state manifold, one can derive all related geometrical entities \cite{Nakahara03}.
We will focus in particular on the geodesics, i.e., the paths of minimal length, and on the coordinate-independent curvature of the underlying space given by the Ricci scalar.

\begin{figure}[t]
    \centering
\includegraphics{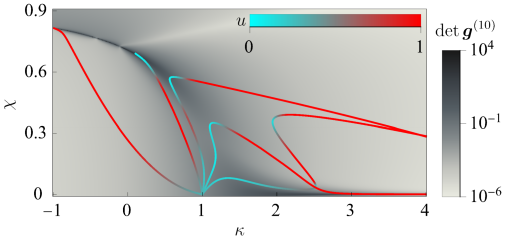}
    \caption{Geodesics (colored curves) on the background of the metric tensor determinant (shades of grey) for the ground-state manifold of the model introduced in Sec.\,\ref{sec:LMG} ($N=10$) with control parameters $\llambda\equiv(\kappa,\chi)$.
    The geodesics are obtained from the Cauchy boundary condition with initial values ${\llambda=(1,0)}$ and $\frac{\d}{\d\tau}\llambda$ defining various exit directions in the $\kappa\times\chi$ plane. Several general features are demonstrated: (i)~Various geodesics can intersect, indicating multiple solutions of the geodesic equation with the Dirichlet boundary condition. (ii)~The ordinary speed $u$ in $(\kappa,\chi)$, expressed in a~relative scale by the varying color of individual curves, is reduced in regions with large metric tensor determinant. (iii)~Some geodesics have a~tendency to be repelled or attracted from/to the DP at $(\kappa,\chi)=(-\frac{1}{9},\sqrt{\frac{10}{19}})$.}
    \label{fig:geods_illustration}
\end{figure}

Let a~function $\llambda(\tau)$, depending on a~single real parameter $\tau\in[\tau_{\rm i},\tau_{\rm f}]$, define a~path ${\cal P}$ between some initial and final points $\llambda_{\rm i}=\llambda(\tau_{\rm i})$ and $\llambda_{\rm f}=\llambda(\tau_{\rm f})$ in the parameter space ${\mathbb\Omega}$.
The geometric length of this curve is given by
\begin{equation}
  \ell^{(N)}_{\cal P}=\int_{{\cal P}}\d\ell = \int_{\tau_{\rm i}}^{\tau_{\rm f}}\sqrt{g^{(N)}_{\mu\nu}\bigl(\llambda(\tau)\bigr)\frac{\d\lambda_{\mu}}{\d\tau}\frac{\d\lambda_{\nu}}{\d\tau}}\ \d\tau
  \label{eq:length}.
\end{equation}
Applying the variational principle to $\ell^{(N)}_{{\cal P}}$, one obtains the following equation for the geodesic,
\begin{eqnarray}
    && \frac{\d^2 \lambda_\mu}{\d\tau^2} + \Gamma^{(N)}_{\mu\sigma\rho} \frac{\d\lambda_\sigma}{\d\tau} \frac{\d\lambda_\rho}{\d\tau} = 0
    \label{eq:geodesic},
\\
   && \Gamma^{(N)}_{\alpha\beta\gamma} = \frac{1}{2}\,{g^{(N)}_{\alpha \mu}}^{-1}\,\bigl(\partial_{\gamma}g^{(N)}_{\mu \beta}+\partial_{\beta}g^{(N)}_{\gamma\mu}-\partial_{\mu}g^{(N)}_{\beta\gamma}\bigr)
   \label{eq:christ},
\end{eqnarray}
where $\Gamma^{(N)}_{\alpha\beta\gamma}$ are the Christoffel symbols defined with the aid of elements of the inverse metric tensor ${\bm{g}^{(N)}}^{-1}$ \cite{Nakahara03} (here and below we again hide all dependencies on $\llambda$).
Note that any solution to this equation always yields a~constant value of the quantity ${v=\frac{\d}{\d\tau}\ell}$, given by the square root in Eq.\,\eqref{eq:length}, which is the geometric speed on the manifold if the parameter $\tau$ coincides with time $t$.
As a~consequence, the ordinary speed
\begin{equation}
    u=\frac{\d}{\d\tau}s=\sqrt{\delta_{\mu\nu}\frac{\d\lambda_{\mu}}{\d\tau}\frac{\d\lambda_{\nu}}{\d\tau}}
    \label{eq:plainspeed}
\end{equation}
in the space of parameters $\llambda$ (with $s$ denoting the ordinary distance in this space assuming the Cartesian metric) decreases in the regions where the metric tensor has large components.
This is illustrated (for a~model introduced below) in Fig.\,\ref{fig:geods_illustration}.
To visualize variations of the overall geometric scale in the parameter space, we plot in Fig.\,\ref{fig:geods_illustration} and some figures below the metric tensor determinant ${\rm det}\,\bm{g}^{(N)}(\llambda)$, which represents the ratio between the geometric and ordinary volumes of an infinitesimal cell in the parameter space at a~given point $\llambda$.

The geodesic equation \eqref{eq:geodesic} can be solved with the Dirichlet or Cauchy boundary condition.
In the Dirichlet case, which is consistent with the original intention for the use of \eqref{eq:geodesic}, both the initial and final values of $\llambda$ are fixed (defining points $\llambda_{\rm i}$ and $\llambda_{\rm f}$) and the solution is supposed to represent the shortest path between these points.
In the Cauchy case, the fixed quantities are the initial values of $\llambda$ and $\frac{\d}{\d\tau}\llambda$, so the solution represents the geodesic emitted from the initial point $\llambda_{\rm i}$ in the selected direction.
This mode of solution of the geodesic equation is computationally simpler.
One may expect that the Dirichlet and Cauchy approaches yield equivalent results (as in classical mechanics formulated in terms of the action integral and the Lagrange equations), but this is not generally true.
It turns out that solutions of the Cauchy problem corresponding to two different initial conditions can cross each other, hence the Dirichlet problem can have multiple solutions.
This situation is exemplified in Fig.\,\ref{fig:geods_illustration}.

The scalar curvature (also called the Ricci scalar) 
\begin{eqnarray}
    \Ric^{(N)}={g^{(N)}_{\sigma\nu}}^{-1}\bigl(&&\partial_\mu\Gamma^{(N)}_{\mu\nu\sigma} - \partial_\nu\Gamma^{(N)}_{\mu\mu\sigma}
    \nonumber\\
    && +\,\Gamma^{(N)}_{\mu\mu\lambda}\Gamma^{(N)}_{\lambda\nu\sigma} - \Gamma^{(N)}_{\mu\nu\lambda}\Gamma^{(N)}_{\lambda\mu\sigma}\bigr),
    \label{eq:ricci}
\end{eqnarray}
describes the local geometry of the Riemannian manifold at a~given point $\llambda$ in a~way independent of a~particular parametrization \cite{Nakahara03}.
For two-dimensional manifolds, the Ricci scalar is related to the Gaussian curvature $K=\kappa_1\kappa_2$ (where $\kappa_1$ and $\kappa_2$ are the principal curvatures) through $\Ric=2K$. 
So the cases $\Ric>0$, $\Ric<0$, and $\Ric=0$ distinguish three principal types of invariant local geometry (elliptic, hyperbolic, and cylindrical or flat, respectively).
For three- and higher-dimensional manifolds, however, the Ricci scalar does not suffice to fully determine the type of invariant local geometry.

We see that the formulas for Christoffel symbols and for quantities characterizing invariant geometry, cf.~Eqs.\,\eqref{eq:christ} and \eqref{eq:ricci}, contain components of the inverse metric tensor~${\bm{g}^{(N)}}^{-1}$.
Thus no geometry is defined when the metric tensor~$\bm{g}^{(N)}$ is degenerate (has zero determinant) and its inversion is impossible.
However, even if the metric tensor is not at exact degeneracy but only close to it, the resulting geometry is not robustly defined.
To demonstrate this, let us focus on the case with ${D=2}$. 
A general nearly degenerate metric tensor in two dimensions reads as
\begin{equation}
    \bm{g}^{(N)}=\left(\begin{array}{cc}a^2&ab\\ab&b^2\end{array}\right)
    +\left(\begin{array}{cc}\alpha&\gamma\\\gamma&\beta\end{array}\right),
    \label{eq:degg}
\end{equation}
where ${a,b,\alpha,\gamma\in\mathbb{R}}$ are arbitrary parameters.
The first term in Eq.\,\eqref{eq:degg} is exactly degenerate, while the second term is assumed to be only a~tiny correction proportional to a~small number $\epsilon$. 
The corresponding inverse metric tensor is
\begin{eqnarray}
\label{eq:degginv}
    {\bm{g}^{(N)}}^{-1}=&&\frac{1}{{\rm det\,}\bm{g}^{(N)}}
    \left(\begin{array}{cc}b^2+\alpha&-ab-\gamma\\-ab-\gamma&a^2+\beta\end{array}\right),
    \\
    &&\ {\rm det\,}\bm{g}^{(N)}=a^2\beta+b^2\alpha-2ab\gamma+\alpha\beta-\gamma^2.
    \nonumber
\end{eqnarray}
Elements of this matrix are large, in the leading order proportional to $1/\epsilon$.
Therefore, even a~very small change of the metric tensor near degeneracy produces huge changes in the geometric quantities, which makes the construction of geometry unstable and practically impossible for numerical computations.

As the last remark, let us emphasize the different roles of invariant (coordinate-independent) and non-invariant (coordinate-dependent) geometric properties in applications to specific quantum models.
The invariant geometry unveils the universal properties of the ground-state (or any excited-state) manifold.
In contrast, the coordinate-dependent geometry is important for problems with predefined parameter space, e.g., in quantum-state driving. 
In the following, we sometimes use non-invariant concepts, such as local diagonalization of the metric tensor and the metric determinant, which can provide information about the interplay between the coordinates of choice and the underlying manifold.

\subsection{Geometric implications of DPs}
\label{sec:DP}

When the energy gap between the ground state and the first excited state closes, the geometry exhibits singular behavior.   
For degenerate ground states, the definition of distance in Eq.\,\eqref{eq:distanc} becomes problematic and, as Eq.\,\eqref{eq:geopert} suggests, the metric tensor elements may diverge as we approach the degeneracy.
Although, as shown below, this divergence is not present in all coordinate systems, the geometry near degeneracies needs special treatment. 

Degeneracies can occur in any region of the parameter space $\mathbb{\Omega}$, forming subsets of lower dimension. 
For finite systems with ${D\geq 2}$ parameters, these subsets likely have dimension ${D-2}$. 
In the simplest case of ${D=2}$, they form isolated points known as diabolic points~\cite{Berry84,Wilczek88}.
In some cases, DPs represent precursors of QPTs that emerge in the limit of infinite system size. 
This happens if an increasing number of DPs concentrates with growing $N$ along a~1-dimensional subset of $\mathbb{\Omega}$, which in the ${N\to\infty}$ limit becomes the QPT borderline.
We will see an example of such behavior in Sec.\,~\ref{sec:LMG}.
However, the existence of DPs is, in fact, neither necessary nor sufficient condition for the creation of a~QPT.

The essential model for a~pairwise crossing of levels in the DP is a~two-level model with Hamiltonian
\begin{equation}
\hat{\mathsf{H}}=c-
      \underbrace{r\sin\theta\cos\phi}_{x}\,\hat{\sigma}_x-\underbrace{r\sin\theta\sin\phi}_{y}\,\hat{\sigma}_y-\underbrace{r\cos\theta}_{z}\,\hat{\sigma}_z,
  \label{eq:ham2}
\end{equation}
where $\hat{\sigma}_x,\hat{\sigma}_y,\hat{\sigma}_z$ are Pauli matrices, $x,y,z\in\mathbb{R}$ stand for dimensionless control parameters, and $c$ is an arbitrary energy shift.
The DP is located at the origin ${(x,y,z)=(0,0,0)}$.
After the transformation of the Cartesian coordinates $(x,y,z)$ to spherical coordinates $(r,\theta,\phi)$, it becomes obvious that variations of $r$ in Eq.\,\eqref{eq:ham2} do not induce any changes of eigenstates.
Therefore, the ${D=3}$ metric tensor matrix ${\{\mathsf{g}_{\mu\nu}\}\equiv\boldsymbol{\mathsf{g}}}$ associated with Hamiltonian~\eqref{eq:ham2} in Cartesian coordinates $(x,y,z)$ is degenerate at any parameter point, having always one null eigenvalue corresponding to the radial direction.
It is well known that an essential parametrization of the two-level problem has dimension ${D=2}$ \cite{Berry84}.
Nevertheless, the conclusion that the direction straight to/from the DP in a~general parameter space does not generate any geometrical distance will be shown to have remarkable consequences.

\begin{figure}
    \centering
    \includegraphics[width=0.75\linewidth]{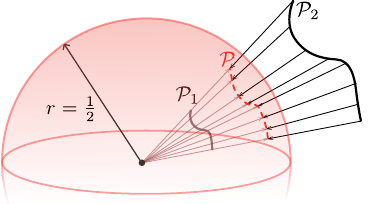}
    \caption{In the two-level system, Eq.\,\eqref{eq:ham2}, the geometric distance along any path in the parameter space (full lines~$\mathcal P_1$ and~$\mathcal P_2$) equals the length of its projection (dashed line $\mathcal P$) to the sphere with radius ${r=\frac{1}{2}}$. The geometric distance along the path close to (far from) the DP is thus enlarged (compressed) relative to the respective Cartesian distances.}
    \label{fig:distance-on-sphere}
\end{figure}

Components $\mathsf{g}_{\mu\nu}$ of the two-level metric tensor $\boldsymbol{\mathsf{g}}$ in Cartesian coordinates $(x,y,z)$ diverge as the ordinary distance ${r=\sqrt{x^2\!+\!y^2\!+\!z^2}}$ from the DP goes to zero (but if approaching the DP along the $x$, $y$ or $z$ axis, the corresponding diagonal element $\mathsf{g}_{xx}$, $\mathsf{g}_{yy}$ or $\mathsf{g}_{zz}$ is zero). 
This divergence is regularized by the transformation to spherical coordinates $(r,\theta,\phi)$, where the metric tensor becomes diagonal with ${\mathsf{g}_{rr}=0}$, ${\mathsf{g}_{\theta\theta}=\frac{1}{4}}$ and ${\mathsf{g}_{\phi\phi}=\frac{1}{4}\sin^2\theta}$.
So the squared distance element reads as
\begin{equation}
\mathrm{d}\ell^2 = \frac{1}{4}\bigl(\mathrm{d}\theta^2 + \sin^2\theta\,\mathrm{d}\phi^2\bigr),
  \label{eq:gdia}
\end{equation}
which describes geometry of the 2-sphere with radius ${r=\frac{1}{2}}$ \cite{Matus23b}. 
This is in accord with the constant positive Ricci scalar $\Ric=8$ associated with Hamiltonian~\eqref{eq:ham2} in arbitrary nonredundant ${D=2}$ coordinates.

Formula~\eqref{eq:gdia} implies that the geometric length \eqref{eq:length} of an arbitrary path in the parameter space $(x,y,z)$ is determined by its radial projection from the centre ${r=0}$ to the sphere ${r=\frac{1}{2}}$. 
This is sketched in Fig. \ref{fig:distance-on-sphere}. 
Consequently, the geometric length $\ell_{\mathcal{P}}$ of paths very close to the DP is strongly enlarged in comparison to their ordinary length ${s_{\mathcal{P}}=\int_{\mathcal{P}}ds}$ in the parameter space. 
The magnification becomes asymptotically large in an infinitesimal vicinity of the DP. 
A trivial consequence of the spherical geometry is that the geometric length of any closed path, which encircles the DP and together with it defines a~single plane in the $(x,y,z)$ space, is given by ${\ell_{\circ}=\pi}$. 

To extend these insights to multi-level systems, we need to approximate the Hamiltonian near the DP by a~two-level Hamiltonian involving only the pair of levels that cross at the DP.
At first, the full Hamiltonian at parameter values $\llambda$ near the DP point $\llambda_{\rm d}$ is represented by a~suitably truncated Taylor expansion
\begin{eqnarray}
    \hat{\mathsf{H}}_{{\rm d}}^{(N)}(\Delta\llambda)=&&\hat{H}^{(N)}(\llambda_{\rm d})+\partial_\mu \hat{H}^{(N)}(\llambda_{\rm d})\Delta\lambda_\mu
    \nonumber\\
    &&\qquad+\frac{1}{2}\partial_\mu\partial_\nu\hat{H}^{(N)}(\llambda_{\rm d})\Delta\lambda_\mu\Delta\lambda_\nu+\dots,
    \qquad
\end{eqnarray}
where ${\Delta\llambda=\llambda-\llambda_{\rm d}}$. 
The two-level approximation is then obtained from matrix elements of this expansion in arbitrary states $\ket{0}$ and $\ket{1}$ forming an orthonormal basis of the degenerate ground-state subspace
at ${\llambda=\llambda_{\rm d}}$.
This leads to the description of the ground state and the first excited state near the DP by an effective two-level Hamiltonian \eqref{eq:ham2} with
\begin{equation}
\hspace{-3mm}
\begin{array}{l}
    c(\Delta\llambda)\!=\!\frac{\matr{0}{\hat{\mathsf{H}}_{{\rm d}}^{(N)}(\Delta\llambda)}{0}+\matr{1}{\hat{\mathsf{H}}_{{\rm d}}^{(N)}(\Delta\llambda)}{1}}{2}\!=\!E_{\rm d}\!+\!{\cal O}(\Delta\llambda),
    \\
    x(\Delta\llambda)\!=\!\frac{\matr{1}{\hat{\mathsf{H}}_{{\rm d}}^{(N)}(\Delta\llambda)}{0}+\matr{0}{\hat{\mathsf{H}}_{{\rm d}}^{(N)}(\Delta\llambda)}{1}}{2}\!=\!{\cal O}(\Delta\llambda),
    \\
    y(\Delta\llambda)\!=\!\frac{\matr{1}{\hat{\mathsf{H}}_{{\rm d}}^{(N)}(\Delta\llambda)}{0}-\matr{0}{\hat{\mathsf{H}}_{{\rm d}}^{(N)}(\Delta\llambda)}{1}}{2\iu}\!=\!{\cal O}(\Delta\llambda),
    \\
    z(\Delta\llambda)\!=\!\frac{\matr{0}{\hat{\mathsf{H}}_{{\rm d}}^{(N)}(\Delta\llambda)}{0}-\matr{1}{\hat{\mathsf{H}}_{{\rm d}}^{(N)}(\Delta\llambda)}{1}}{2}\!=\!{\cal O}(\Delta\llambda),
\end{array}
\label{eq:xyz}
\end{equation}
where $E_{\rm d}$ is the energy of the DP and the symbol ${\cal O}(\Delta\llambda)$ represents polynomial expressions in $\Delta\lambda_\mu$ whose lowest order is generically equal to unity.
The above equation defines a~surface in the $(x,y,z)$ space, which for ${\Delta\llambda=0}$ passes the DP at the origin. 
In an infinitesimal vicinity of the DP, the surface can be approximated by the tangent plane given by the linear terms in Eq.\,\eqref{eq:xyz}, but due to the higher-order terms the surface is generically curved.

To avoid redundant parametrizations of the two-level problem, we restrict the discussion of the near-DP geometry to systems with ${D=2}$, so ${\llambda\equiv(\lambda_1,\lambda_2)}$ (moreover, for ${D>2}$ the degeneracies most likely do not form isolated points).
The mapping~\eqref{eq:xyz} together with the metric~\eqref{eq:gdia} give rise to a~two-level approximation $\boldsymbol{\mathsf{g}}^{(N)}_{{\rm d}}(\Delta\llambda)$ of the actual ground-state metric tensor $\bm{g}^{(N)}(\llambda)$ in a~vicinity of the DP.

The presence of both linear and nonlinear terms in variable $\Delta\lambda_\mu$ on the right-hand side of Eq.\,\eqref{eq:xyz}, which can be assumed in a~typical situation, has important implications.
The nonlinear terms lead to a~generically nontrivial dependence of the approximate metric tensor~$\boldsymbol{\mathsf{g}}^{(N)}_{{\rm d}}(\Delta\llambda)$ and the resulting geometry on $\Delta\llambda$ close to the DP.
This dependence, in general, lacks symmetry with respect to rotations around the DP.
On the other hand, the linear terms prevail very close to the DP.
In this domain, the approximate metric tensor $\boldsymbol{\mathsf{g}}^{(N)}_{{\rm d}}(\Delta\llambda)$ nearly coincides with a~cut through the simple two-level metric tensor $\boldsymbol{\mathsf{g}}(x,y,z)$ by the tangent plane to the surface \eqref{eq:xyz} at the DP, which makes the approximate metric tensor nearly degenerate due to the almost vanishing distance element along the radial direction to/from the DP.
Therefore, according to Eqs.\,\eqref{eq:degg} and \eqref{eq:degginv}, the geometry gets very unstable for very small values of $\Delta\lambda_\mu$ and collapses in the limit $\Delta\llambda\to 0$.
Note that the singular geometry and attraction of some geodesics to the DP (due to the vanishing radial distance element) make DPs, in some sense, analogous to black holes.

In the vicinity of the DP, we introduce new polar coordinates ${\mathcal{R}\in[0,\infty)}$ and $\mathit{\Theta}\in[0,2\pi)$, so that
\begin{equation}
    (\Delta\lambda_1,\Delta\lambda_2)=(\mathcal{R}\cos\mathit{\Theta},\mathcal{R}\sin\mathit{\Theta}).
    \label{eq:pola}
\end{equation}
These are analogous to $r$ and $\theta$ from Eq.\,\eqref{eq:ham2} with any fixed angles $\phi$ and ${\phi+\pi}$ (the present angle $\mathit{\Theta}$, in contrast to $\theta$, covers the whole circle around the DP).
The polar coordinates $(\mathcal{R},\mathit{\Theta})$ regularize the DP divergence of the metric tensor components, which can, in generic cases, be expected in the original coordinates $(\Delta\lambda_1,\Delta\lambda_2)$.
Despite the lack of the rotational symmetry of $\boldsymbol{\mathsf{g}}^{(N)}_{{\rm d}}(\Delta\llambda)$ around the DP, the integration in angle $\mathit{\Theta}$ leads to an analogous result as in the actual two-level case. 
Indeed, as follows from the above considerations (for another perspective see Ref.\,\cite{Mailybaev18}), the geometric length $\ell_{\circ}$ of a~small circle ${\mathcal{R}=\mathrm{const}}$ around the DP is given by
\begin{equation}
    \ell_{\circ}=\pi+f(\mathcal{R}),
    \label{eq:circ}
\end{equation}
where $f(\mathcal{R})$ is a~semipositive function limiting to zero for ${\mathcal{R}\to 0}$.

Although the transition to polar coordinates \eqref{eq:pola} removes the DP divergence of the metric tensor components, the form $\boldsymbol{\mathsf{g}}^{(N)}_{{\rm d}}(0)$ right at the DP is not defined as the ${\mathcal{R}\to 0}$ limit of the metric tensor depends on $\mathit{\Theta}$.
To address this problem, it is customary to reinterpret the DP as a~boundary in the parameter space. 
This is achieved by transforming the polar coordinate according to ${\mathcal{R}\mapsto\mathcal{R}+\mathcal{R}_0}$, where the shift constant $\mathcal{R}_0>0$ pushes the DP away from the origin, associating it with a~circular boundary in the parameter space. 
The natural choice of the shift is $\mathcal{R}_0=\frac{1}{2}$, which reproduces the circumference in Eq.\,\eqref{eq:circ} for ${\mathcal{R}\to 0}$.
This transformation is useful not only for visualisation purposes but also as a~tool for effective analyses of geodesics and other geometrical features of the ground-state manifold near the DP.

Finally, we need to stress that the two-level approximation based on Eq.\,\eqref{eq:xyz} should not be overvalued.
This method roughly captures the near-DP behavior of individual metric tensor components and yields important qualitative insight into the geometrical implications of DPs.
However, it does not provide reasonable estimates of invariant geometrical quantities in the vicinity of the DP.
One reason is that these quantities depend on the metric-tensor derivatives, which are usually not well reproduced by the two-level approximation. 
Another reason follows from the convergence of the metric tensor to a~nearly degenerate form as ${\mathcal{R}\to 0}$, which results in totally unpredictable geometry very close to the DP. 
These issues will be illustrated by a~concrete example in Sec.\,\ref{sec:DPboson}.

\subsection{Geometric implications of QPTs}
\label{sec:QPT}

As mentioned above, degeneracies of the ground and first-excited states in finite systems with $D$ real parameters usually form ${(D\!-\!2)}$-dimensional subsets of the parameter domain.
However, as for some systems the repulsion of levels weakens with increasing size, the degeneracies in the infinite size limit can constitute even ${(D\!-\!1)}$-dimensional subsets---the emergent QPT borderlines forming the ground-state phase separatrices \cite{Sachdev99,Carr11}.
When $\llambda$ passes a~point $\llambda_{\rm c}$ on the phase separatrix, the ground-state wave function $\ket{\psi_0(\llambda)}$ varies in a~nonanalytic way.
A discontinuous (first-order) QPT represents a~sudden jump between two forms of the wave function, which correspond to the two levels crossing at $\llambda_{\rm c}$. 
Continuous QPTs represent continuous dependencies of $\ket{\psi_0(\llambda)}$ with only the $k$th derivative ($k>0$) discontinuous (for the $k$th order QPT) or infinite (for a~continuous QPT without Ehrenfest classification).
These effects are usually connected with slowly developing (with increasing size) degeneracies of a~larger number of states.

What is the impact of QPTs on geometry? 
First, it is evident that the metric tensor cannot be defined at the {\em first-order\/} QPT. 
Indeed, if proceeding along the first-order phase separatrix, the degeneracy disables the definition of distance, while in the direction perpendicular to the separatrix, the derivatives of eigenstates involved in Eq.\,\eqref{eq:geo} do not exist because of the jump of the wave function at the critical point.
Assuming that the transition through the QPT is governed by a~single real parameter~$\lambda$ such that the critical point appears at a~certain value~$\lambda_{\rm c}$, we can see that the square root of the intensive metric tensor~$\sqrt{g_{\lambda\lambda}(\lambda)}$ has a~singularity of the $\delta$-function type at ${\lambda=\lambda_{\rm c}}$.
This singularity is irremovable by any conceivable reparametrization of the Hamiltonian.
One can imagine that a~first-order QPT tears the ground-state manifold into multiple disconnected parts.
This fragmentation of the parameter space makes it impossible for geodesics to traverse the QPT borderline from one ground-state phase to the other \cite{Kumar12}. 

The situation is, however, different for a~{\em continuous\/} QPT.
In this case, assuming again the single $\lambda$ parameter governing the transition through the QPT, the dependence $\sqrt{g_{\lambda\lambda}(\lambda)}$ has a~non-analyticity (divergence, discontinuity, or discontinuity in the first or higher derivatives) at ${\lambda=\lambda_{\rm c}}$, but this singularity is weaker than of the $\delta$-function type.
For continuous QPTs, there always exists a~possibility to reparametrize the Hamiltonian so that the nonanalyticity of $\ket{\psi_0(\lambda)}$ and $g_{\lambda\lambda}(\lambda)$ gets washed out.
As an example, we can transform $\lambda$ using the formula
\begin{equation}
    \lambda=\lambda_{\rm c}+\xi\,e^{-(a/\xi)^2},
\label{eq:smoothen}
\end{equation}
where $\xi$ is a~new parameter defined in a~subset of the real axis containing zero and $a$ is a~suitable positive constant.
It is clear that ${\lambda=\lambda_{\rm c}}$ for ${\xi=0}$, while for ${|\xi|\gg a}$ we get ${\lambda\approx\lambda_{\rm c}+\xi}$.
The function ~$\lambda(\xi)$ is smooth (infinitely differentiable) but non-analytic at ${\xi=0}$, with all derivatives equal to zero, which means that after the mapping~${\lambda\mapsto\xi}$ we get ${\frac{\d^k}{\d\xi^k}\ket{\psi_0(\xi)}\bigr|_{\xi=0}\!=0}$ for all $k\!=\!{1,2,\dots}$. 
So any discontinuity of the wave function derivative at the critical point is gone, but ${g_{\xi\xi}=0}$ at ${\xi=0}$.
The same conclusion also holds for continuous QPTs with divergent derivatives (no Ehrenfest classification) if the divergence is just of an algebraic type (since then, it is always beaten by the exponential vanishing of the respective derivative of~$\lambda$).

\begin{figure}
    \includegraphics[width=\linewidth]{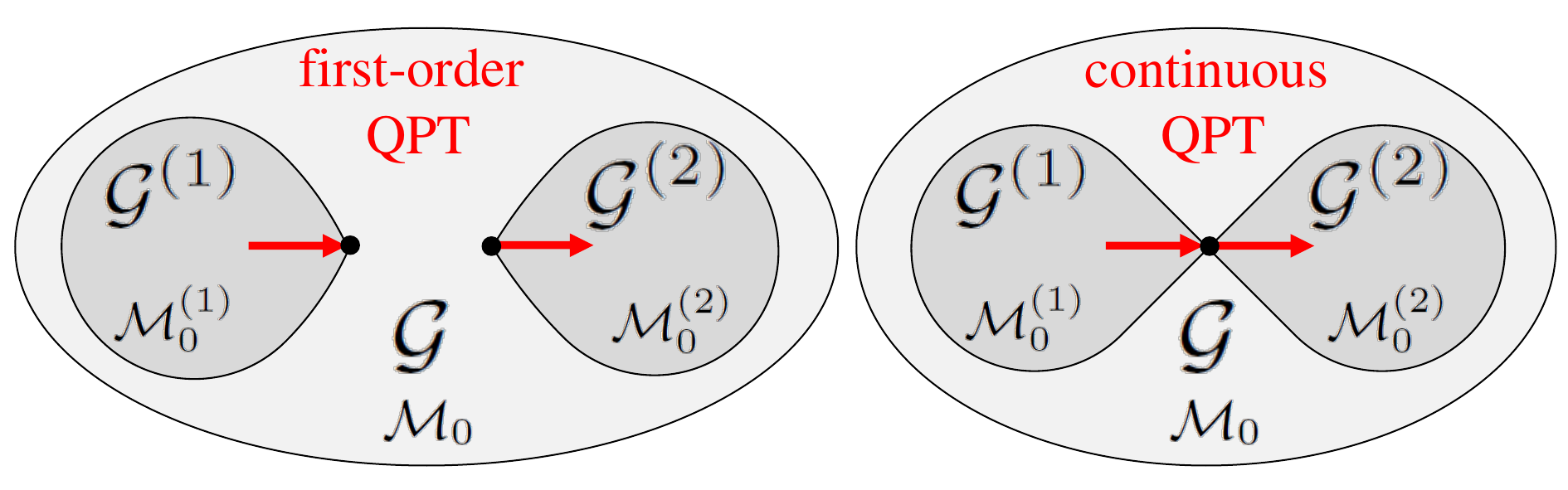}
    \caption{A schematic layout of geometries involved in general QPTs. Manifolds ${\cal M}_0^{(1)}$ and ${\cal M}_0^{(2)}$ of ground states on both sides of a~first-order QPT are unconnected and may constitute incompatible invariant geometries ${\cal G}^{(1)}$ and ${\cal G}^{(2)}$. In contrast, ground-state manifolds for a~continuous QPT are joined, so their geometries must be unifiable. The existence of an enveloping geometry ${\cal G}$ associated with the full ground-state manifold ${\cal M}_0$ of the model is assumed.}
    \label{fig:geophases}
\end{figure}

This means that the continuous QPT, in contrast to the first-order QPT, can be removed by a~suitable transformation to some new coordinates.
However, this comes at a~cost of a~singular geometry with a~vanishing metric tensor at the critical point. 
Our geometric view of general first-order and continuous QPTs is schematically sketched in Fig.\,\ref{fig:geophases}.

Besides geometric implications of the QPT borderlines, one can also study geometric properties of individual ground-state phases, i.e., the geometry of the $D$-dimensional subdomains of the parameter space $\mathbb{\Omega}$ on both sides of the phase separatrix.
Are these quantum phases associated with invariant geometries that define some fundamental geometrical objects underlying the respective ground-state manifolds?
Neat examples of such phase-specific geometries were presented in Ref.\,\cite{Kolodrubetz13}.
Although we will demonstrate (by a~particular example in Sec.\,\ref{sec:geomodel}) that the geometry of the actual ground-state manifold within a~single quantum phase can be rather complicated, the question still lasts whether the embedding to a~space with simpler geometry cannot be achieved via a~suitable higher-$D$ generalization of the model.


\section{A bosonic model}
\label{sec:multi_qubit}

\subsection{General finite-dimensional bosonic models and HB approximation}
\label{sec:genbos}

In this section, we describe a~specific boson model that will later illustrate the general statements of the previous section.
We start by introducing a~wider class of interacting boson models, to which our results can be directly generalized, namely the models that conserve the total number of bosons $N$ and carry a~finite number of degrees of freedom $f$.
Such models are commonly used in the description of collective dynamics of many-body systems such as molecules, atomic nuclei, Bose-Einstein condensates, or cold atoms in optical lattices \cite{Iachello87,Iachello95,Jaksch98,Kawaguchi12}.
Each boson in these models is associated with a~single particle Hilbert space of finite dimension ${f+1}$, interpreted as the number of boson types (bosonic \lq\lq flavours\rq\rq\ and/or \lq\lq colors\rq\rq). 
The number of degrees of freedom is $f$ (one less than the boson type number because of the conservation of $N$), and the dimension of the total Hilbert space for $N$ bosons is ${d=(N\!+\!f)!/(N!f!)}$.

Let the boson of $k$th type, with ${k=0,1,\dots,f}$, be created and annihilated by operators $\hat b_k^\dagger$ and $\hat b_k$, respectively (numbering of bosons is arbitrary).
If the number of bosons is conserved and fixed at the value $N$, a~general Hamiltonian can be written as
\begin{equation}
\hat{H}^{(N)}(\llambda)\!=\!\!
\sum_{kl}\!u_{kl}(\llambda)\,\hat{b}_k^\dagger\hat{b}_l
+\!\!\!\sum_{klmn}\!\!\frac{v_{klmn}(\llambda)}{N}\,\hat{b}_k^\dagger\hat{b}_l^\dagger\hat{b}_m\hat{b}_n
\!+\!\dots,
\label{eq:hamib}
\end{equation}
where $u_{kl}$ and $v_{klmn}/N$ stand for one- and two-body interaction strengths depending on some control parameters $\llambda$, and the dots indicate possible three-body and higher interaction terms.
The $1/N$ scaling of the two-body strengths ensures that the two-body part of the Hamiltonian (its expectation values in individual eigenstates of the full Hamiltonian) remains in the same proportion with the one-body part as the number of bosons~$N$ increases.
For the same reason, the $k$-body strengths would be scaled by $1/N^{k-1}$.

A common tool for the description of the ground state of Hamiltonian \eqref{eq:hamib} is the mean field Hartree-Bose (HB) approximation \cite{Blaizot86}, which makes use of the boson condensate states 
\begin{eqnarray}
    \ket{\psi_{\rm HB}^{(N)}(\aalpha)} &=& 
    \frac{1}{\sqrt{N!}}\bigl[\hat{B}(\aalpha)^\dagger\bigr]^N\ket{0},
    \label{eq:hbstates}\\
    \hat{B}(\aalpha)^\dagger &=&
    \frac{1}{\sqrt{S(\aalpha)}}\biggl(\hat{b}_0^\dagger + \sum_{k=1}^{f} \underbrace{\rho_k e^{\iu\phi_k}}_{\alpha_k}\hat{b}_k^\dagger\biggr),
    \label{eq:condensate}
 \end{eqnarray}
where $\ket{0}$ stands for the vacuum state and the operator $B^\dag$ creates bosons to the condensate single-particle state determined by complex parameters ${\aalpha=(\alpha_0,\dots,\alpha_{f})}$ with absolute values ${\rho_k\in[0,\infty)}$ and phases ${\phi_k\in[0,2\pi)}$ that quantify amplitudes of individual bosons $\hat{b}_k^\dagger$ in $\hat{B}^\dag$.
Note that for the ${k=0}$ boson, we set ${\phi_0=0}$ (which is possible due to the freedom of choosing the overall phase factor of the whole state) and~${\rho_0=1}$.
The normalization is ensured by the pre\-factor with ${S=1+\sum_{k=1}^{f}\rho_k^2}$.

The HB approximation of the ground state is obtained by minimizing the finite-$N$ intensive energy functional
\begin{equation}
    {\cal E}^{(N)}(\aalpha,\llambda)=\frac{1}{N}
    \bigl\langle\psi_{\rm HB}^{(N)}(\aalpha)\bigl|\hat{H}^{(N)}(\llambda)\bigr|\psi_{\rm HB}^{(N)}(\aalpha)\bigr\rangle
\label{eq:Emin}
\end{equation}
in variables $\aalpha$ and by performing the ${N\to\infty}$ limit.
Let the minimum for a~given $\llambda$ and $N$ is found at $\aalpha_{\rm min}^{(N)}(\llambda)$.
Then the ${N\to\infty}$ limit of ${\cal E}^{(N)}(\aalpha_{\rm min}^{(N)}(\llambda),\llambda)$ coincides with the same limit of the actual intensive ground-state energy $E_0^{(N)}(\llambda)/N$.
Note that degeneracies of the ground state at some particular values of $\llambda$ are connected with the occurrence of two or more degenerate global minima of Eq.\,\eqref{eq:Emin}. 

\begin{figure}[t!]
    \centering
    \includegraphics{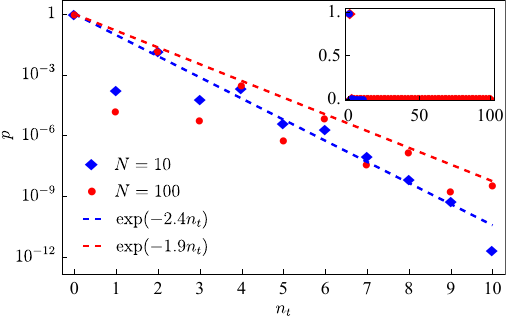}
    \caption{An example of the ground-state decomposition \eqref{eq:exstates} for the ${f=1}$ boson model from Sec.\,\ref{sec:LMG}. Graphs show the dependence of ${|A^{(N)}_{n_t}|^2=|\scal{n_s,n_t}{\psi_0^{(N)}}|^2\equiv p}$ (the probability of finding in the exact ground state the numbers $n_t$ and ${n_s=N-n_t}$ of the $t$ and $s$ bosons, respectively) on $n_t$ for Hamiltonian \eqref{eq:hamibst} at the point $\llambda\equiv(\kappa,\chi)=(-1,0.2)$. This $\llambda$ belongs to the ground-state phase associated in the HB approximation with the $s$-boson condensate. We show the cases with ${N=10}$ and 100. Even-$n_t$ probabilities are fitted by an exponential function (dashed lines), odd-$n_t$ probabilities follow this dependence only for larger $n_t$ and vanish identically at $\chi=0$. The inset shows the same dependencies in the linear scale of $p$ and the full range of ${n_t\in[0,N]}$.} 
    \label{fig:HB_vector_components}    
\end{figure}

Whereas the infinite-size limit of intensive ground-state energy of the bosonic system \eqref{eq:hamib} is correctly evaluated by the HB method, the representation of the ground-state wave function is more subtle.
One naturally expects that the ${N\to\infty}$ limit of $\ket{\psi_0^{(N)}(\llambda)}$ coincides with the same limit of the condensate state $\ket{\psi_{\rm HB}^{(N)}(\aalpha_{\rm min}^{(N)}(\llambda))}$, but this is not precisely true.
Skipping the dependence on $\llambda$, the actual ground state for any~$N$ can be written as
\begin{equation} 
    \ket{\psi^{(N)}_0}=
\!\!\!\!\sum_{n_1,n_2,\dots}
\!\!\!\!
    A^{(N)}_{n_1n_2\dots}
    \frac{[\hat{B}^\dagger]^{N-n_{\perp}}[\hat{B}^\dagger_{1}]^{n_1}[\hat{B}^\dagger_{2}]^{n_2}\dots}{\sqrt{(N\!-\!n_{\perp})!\,n_1!\,n_2!\dots}}\ket{0},
    \label{eq:exstates}
\end{equation}
where operators $\{\hat{B}^\dagger,\hat{B}^\dagger_{1},\hat{B}^\dagger_{2},\dots\}$
create bosons to mutually orthogonal states,  the first of which is identical with the ${\aalpha=\aalpha_{\rm min}^{(N)}(\llambda)}$ condensate state from Eq.\,\eqref{eq:condensate}, and normalized amplitudes $A^{(N)}_{n_1n_2\dots}$ weight components corresponding to numbers ${\{N\!-\!n_{\perp},n_1,n_2,\dots\}}$ of these bosons, with ${n_{\perp}=n_1+n_2+\dots}$.
An example of such a~decomposition of the ground state for the ${f=1}$ model described below is shown in Fig.\,\ref{fig:HB_vector_components}.

The dominant part of Eq.\,\eqref{eq:exstates} comes from the ${n_{\perp}=0}$ term, which represents the full condensate \eqref{eq:hbstates}, but components with finite values of~${n_{\perp}>0}$ are present even in the ${N\to\infty}$ limit and the condensate contribution $\lim_{N\to\infty}|A^{(N)}_{00\dots}|^2$ may be visibly less than unity.
However, since all these ${n_{\perp}>0}$ components contain vanishing (for ${N\to\infty}$) relative populations $n_1/N, n_2/N,\dots$ of the non-condensate states, the wave function \eqref{eq:exstates} can be treated as a~quasi-condensate rather similar to the full condensate.  
Only in this restricted sense, we can say that the HB estimate of the ground-state wave function is valid in the infinite-size limit.

Despite these limitations, the HB method represents a~suitable tool to describe ground-state QPTs in general bosonic systems of the type~\eqref{eq:hamib} \cite{Dieperink80,Feng81,Cejnar07}.
In particular, the QPTs are associated with the values of control parameters $\llambda$ at which the function $\aalpha_{\rm min}^{(N)}(\llambda)$ becomes nonanalytic in the ${N\to\infty}$ limit, either being discontinuous (for the first-order QPT) or having discontinuous or diverging derivatives (for continuous QPTs).
However, the HB approximation suffers from serious drawbacks if used for quantum geometric purposes.
While the HB approximation of the metric tensor is outlined in Appendix~\ref{sec:bosco}, its inaccuracy is manifested in Sec.\,\ref{sec:geomodel}.

\subsection{The \textit{\textbf{f}}=1 boson model}
\label{sec:LMG}

\subsubsection{Hamiltonian and its qubit interpretation}
\label{sec:qubit}

In the following, we will study a~model with $f=1$ belonging to a~wider family of Lipkin-Meshkov-Glick (LMG) models \cite{Lipkin65}.
The bosons ${\hat{b}_0^\dag\equiv\hat{s}^\dag}$ and ${\hat{b}_1^\dag\equiv\hat{t}^\dag}$, respectively, are interpreted as scalar (positive-parity) and pseudoscalar (negative-parity) bosons.
We consider the following Hamiltonian
\begin{align}
\hat{H}^{(N)}&(\kappa,\chi)=E_0(\kappa,\chi)
\label{eq:hamibst}\\
&+\frac{1}{2}\bigl(\hat{t}^\dag\hat{t}-\hat{s}^\dag\hat{s}\,\bigr) 
-\frac{1}{N}\biggl[\chi^2\hat{t}^\dag\hat{t}+\frac{\chi}{2}\bigl(\hat{t}^\dag\hat{s}+\hat{s}^\dag\hat{t}\,\bigr)\biggr]
\nonumber\\
&-\frac{1}{N}\biggl[\frac{\kappa}{4}\, \bigl(\hat{t}^\dag\hat{t}^\dag\hat{s}\hat{s}+\hat{s}^\dag\hat{s}^\dag\hat{t}\hat{t}+2\hat{t}^\dag\hat{s}^\dag\hat{s}\hat{t}\,\bigr)
\nonumber\\
&\qquad\qquad\quad +\chi\,\bigl(\hat{t}^\dag\hat{t}^\dag\hat{s}\hat{t}+\hat{t}^\dag\hat{s}^\dag\hat{t}\hat{t}\,\bigr)
+\chi^2\,\hat{t}^\dag\hat{t}^\dag\hat{t}\hat{t}\,\biggr],
\nonumber
\end{align}
where $\kappa,\chi\in\R$ are two control parameters defining the ${D=2}$ dimensional parameter space $(\kappa,\chi)=(\lambda_1,\lambda_2)=\llambda$ and $E_0$ is an arbitrary energy shift (here set to ${E_0=-\kappa/4}$).
Note that the Hamiltonian and all energies are assumed to be dimensionless, expressed in suitable units not specified here.
The terms on the second line of Eq.\,\eqref{eq:hamibst} are one-body operators, while the terms on the third and fourth lines stand for two-body interactions.
The $1/N$ scaling of the two-body terms ensures that these yield comparable contributions with unscaled one-body terms for any $N$. 
The $1/N$-scaled one-body terms (in square brackets on the second line) fade away for $N\to\infty$.
The specific form \eqref{eq:hamibst} of the Hamiltonian, which is a~special case of the form \eqref{eq:hamib}, was used in Refs.\,\cite{Matus23a,Cejnar23,Matus24}, and we employ it again because of its complex but tractable properties described below.

We see that for ${\chi=0}$ Hamiltonian \eqref{eq:hamibst} conserves the parity defined as ${\hat{\Pi}=(-1)^{\hat{n}_s-\hat{n}_t}}$, where ${\hat{n}_{k}=\hat{b}_{k}^\dag\hat{b}_{k}}$ is the number operator for bosons of the given type, so the system possesses a~discrete symmetry under the cyclic group~$Z_2$.
For ${\chi\neq 0}$, the symmetry is broken.
The presence or absence of the $Z_2$ symmetry has consequences for the nature of ground-state QPTs.

For ${f=1}$, the operators $\hat{b}_k^\dag\hat{b}_{k'}$ with a~fixed value of the total boson number $N$ form the algebra SU(2). 
It can be transformed to the ordinary form through the Schwinger mapping
\begin{equation}
\biggl(\hat{t}^{\dag}\hat{s},\hat{s}^{\dag}\hat{t},\frac{\hat{t}^{\dag}\hat{t}-\hat{s}^{\dag}\hat{s}}{2}\biggr)
\mapsto
\biggl(\hat{J}_+,\hat{J}_-,\hat{J}_{z}\biggr),
\label{eq:schwinger}
\end{equation}
where $\hat{J}_z$ and $\hat{J}_{\pm}=\hat{J}_x\pm\iu\hat{J}_y$ have the standard meaning of angular-momentum (quasispin) operators.
This procedure maps boson states $\ket{n_s,n_t}$ with the boson numbers $n_t=0,\dots,N$ and $n_s=N-n_t$ to the quasispin states $\ket{j,m}$ with $j=N/2$ and $m=-j,\dots,+j$.
The quasispin counterpart of Hamiltonian \eqref{eq:hamibst} reads as
\begin{align}
\hat{H}^{(N)}&(\kappa,\chi)=
\hat{J}_z-\frac{1}{N}\biggl[\kappa\hat{J}_x^2+\chi\bigl\{\hat{J}_x\bigl(\hat{J}_z\!+\!\tfrac{N}{2}\bigr)
\nonumber\\
&+\bigl(\hat{J}_z\!+\!\tfrac{N}{2}\bigr)\hat{J}_x\bigr\}+\chi^2\bigl(\hat{J}_z\!+\!\tfrac{N}{2}\bigr)^2\biggr].
\label{eq:hamspin}
\end{align}
Assuming that the quasispin operators are sums of spin-$\tfrac{1}{2}$ operators assigned to $N$ elementary two-state objects,
\begin{equation}
\hat{J}_{a}=\frac{1}{2}\sum_{i=1}^{N}\hat{\sigma}_{a}^{(i)},
\qquad{a=x,y,z},
\label{J}
\end{equation}
we arrive at a~qubit interpretation of Hamiltonian \eqref{eq:hamspin}.
Its first term just sums energies $\pm\tfrac{\epsilon}{2}$ of individual qubits in up and down states (associated in any order with logical 0 and 1 states), while the terms in square brackets represent interactions between qubits.

With the aid of Eqs.\,\eqref{eq:hamspin} and \eqref{J}, the boson model \eqref{eq:hamibst} is cast as a~model describing $N$ mutually interacting qubits.
Because the interactions include all qubits, the system is said to be fully connected. 
The Hilbert space spanned by states $\ket{j=N/2,m}$ corresponds to an ${(N+1)}$-dimensional subspace of the whole $2^N$-dimensional space of $N$ qubits.
This subspace incorporates states which are fully symmetric under the qubit exchange and can be obtained from all-up or all-down qubit states by repeated action of ladder operators $\hat{J}_-$ or $\hat{J}_+$, respectively.
This $j=N/2$ subspace is invariant under the evolution induced by Hamiltonian \eqref{eq:hamspin}.

\subsubsection{HB approximation and QPTs}
\label{sec:phase}

\begin{figure}[t]
    \centering
    \includegraphics{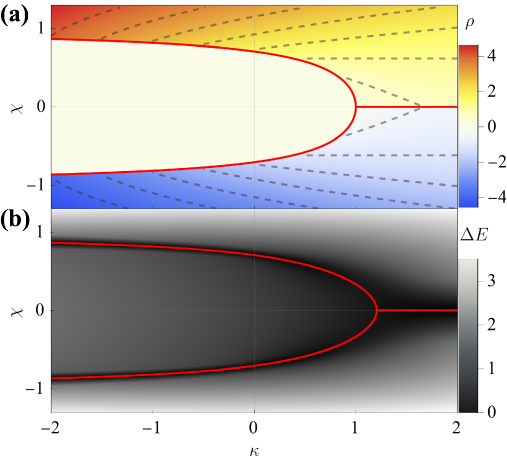}
    \caption{QPTs in the ${f=1}$ boson model \eqref{eq:hamibst}. (a)~Discontinuities in the ${N\to\infty}$ ground-state condensate visualized in a~combined density and contour plot (color scale and dashed lines) of the condensate parameter $\rho(\kappa,\chi)$. The full (red) lines correspond to the QPT phase separatrices from Eq.\,\eqref{eq:qptsep}. (b)~The density plot of the energy gap ${\Delta E^{(N)}(\llambda)}$ for a~finite system with ${N=10}$. The (red) lines mark places with a~locally minimal gap, defining the regions of finite-size QPT precursors, which are slightly shifted with respect to real QPTs in panel~(a).}
    \label{fig:transitionCompare}    
\end{figure}

We can now apply the general condensate technique described in Sec.\,\ref{sec:genbos} to estimate the ground-state energy and wave function of the above Hamiltonian $\hat{H}^{(N)}(\kappa,\chi)$.
The condensate form \eqref{eq:condensate} for the $s$ and $t$ bosons has just a~single complex parameter ${\alpha=\rho e^{\iu\phi}}$.
It turns out that for Hamiltonian \eqref{eq:hamibst} with any boson number~$N$, the minimum of the energy function \eqref{eq:Emin} for all values of $(\kappa,\chi)$ is found either at ${\phi=0}$ or at ${\phi=\pi}$; this reality constraint on the condensate parameters $\alpha_i$ holds for a~general Hamiltonian \eqref{eq:hamib} with real coefficients $u_{kl}$ and $v_{klmn}$.
To simplify the analysis, we will ignore the phase variable, representing the ${\phi=0}$ case by positive values of $\rho$ and the ${\phi=\pi}$ case by negative values of $\rho$. 
So the ground-state estimate is fully determined by a~single real function $\rho^{(N)}_{\rm min}(\kappa,\chi)$.
The $N\to\infty$ limit of this function coincides with the $\rho$ that minimizes the expression 
\begin{equation}
    \frac{-1+\rho^2\bigl[-2\kappa+\rho(\rho-4\chi-2\rho\chi^2)\bigr]}{(1+\rho^2)^2},
    \label{eq:fun}
\end{equation}
and is denoted as $\rho(\kappa,\chi)$.

Nonanalyticities of the function $\rho(\kappa,\chi)$ indicate ground-state QPTs. 
The analysis of Eq.\,\eqref{eq:fun} discloses transitions of four types \cite{Matus23a}:
\begin{equation}
\begin{array}{llll}
    \mbox{QPT-1}_{\pm} &\!\!\!: & \qquad \kappa<1, & \chi=\pm\sqrt{\dfrac{\kappa-1}{\kappa-2}}, \\
    \mbox{QPT-2}       &\!\!\!: & \qquad \kappa=1, & \chi=0,  \\
    \mbox{QPT-1}_0     &\!\!\!: & \qquad \kappa>1, & \chi=0. 
\end{array}
\label{eq:qptsep}
\end{equation}
Here {QPT-1$_+$} and {QPT-1$_-$}, for ${\chi>0}$ and ${\chi<0}$, respectively, as well as {QPT-1$_0$}, for ${\chi=0}$, represent phase transitions of the first order.
In {QPT-1$_{\pm}$, the ground state changes abruptly from the pure $s$ condensate (${\rho=0}$ at lower $\kappa$ or $|\chi|$) to a~mixed $s$ plus $t$ condensate (${\rho\neq 0}$ at higher $\kappa$ or $|\chi|$). 
In {QPT-1$_0$}, the sudden change affects the sign of the $t$-boson admixture in the mixed condensate (${\rho<0}$ for ${\chi<0}$ and ${\rho>0}$ for ${\chi>0}$).
Along the {QPT-1$_0$} line, both ${\rho>0}$ and ${\rho<0}$ solutions coexist, indicating the ground-state degeneracy due to the spontaneously broken parity for ${\kappa>1}$ and ${\chi=0}$.
On the other hand, QPT-2 represents the ground-state phase transition of the second order. 
It occurs at the triple point located at the intersection of three first-order transitions.
Passing this point in any direction leads to a~continuous but undifferentiable change of $\rho$.
The three ground-state forms separated by the QPTs from Eq.\,\eqref{eq:qptsep} are denoted here as phase~I (${\rho=0}$), phase~$\mbox{II}_+$ (${\rho>0}$) and phase~$\mbox{II}_-$ (${\rho<0}$).

The $N\to\infty$ condensate parameter $\rho(\kappa,\chi)$ is depicted in the upper panel of Fig.\,\ref{fig:transitionCompare}.
The plot is consistent with the above analysis.
The lower panel shows the energy gap 
\begin{equation}
  \Delta E^{(N)}(\llambda)=E_1^{(N)}(\llambda)-E_0^{(N)}(\llambda)
\end{equation}
between the exact energies of the first excited state and the ground state for a~moderate-size system (${N=10}$), demonstrating a~relatively good correspondence between the finite-size region with minimal gap and the infinite-size locus of QPTs. 

\subsubsection{Finite-size analysis and DPs}
\label{sec:diabol}

\begin{figure}[t]
    \centering
    \includegraphics{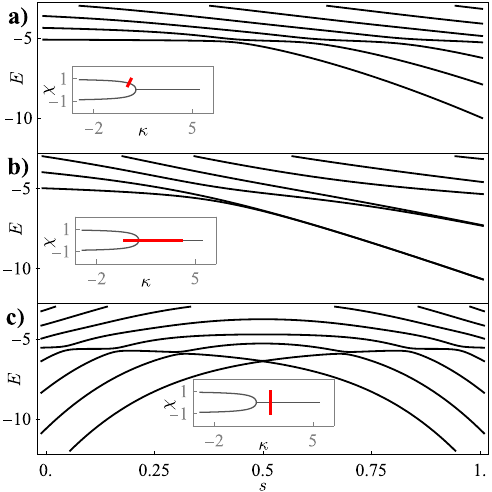}
    \caption{Low-energy spectra of Hamiltonian \eqref{eq:hamibst} with ${N=10}$ along three paths crossing QPT-precursor regions in the $\llambda\equiv(\kappa,\chi)$ plane. The paths across (a) ${\mbox{QPT-1}_{+}}$, (b) ${\mbox{QPT-2}}$ and (c) ${\mbox{QPT-1}_{0}}$ are depicted in the respective insets. They all have the form ${\llambda(\tau)=\llambda_{{\rm i}}+\tau(\llambda_{{\rm f}}-\llambda_{{\rm i}})}$, with parameter ${\tau\in[0,1]}$ measuring the fraction of the path from $\llambda_{{\rm i}}$ to $\llambda_{{\rm f}}$. } 
    \label{fig:N=10_energies}
\end{figure}

The QPTs have sharp effects only in the infinite-size limit, but in Fig.\,\ref{fig:transitionCompare}(b), we saw their clear finite-size precursors at a~finite value of $N$.
Figure~\ref{fig:N=10_energies} shows low-lying parts of energy spectra for ${N=10}$ along three paths in the parameter plane crossing various QPT-precursor regions.
We know that at the first-order QPTs the energy gap $\Delta E^{(N)}(\llambda_{\rm c})$ drops exponentially with $N$ (so ${\Delta E^{(N)}\propto e^{-aN}}$ with a~constant ${a>0}$), while at the second-order QPT it decreases only algebraically (${\Delta E^{(N)}\propto N^{-a}}$ again with ${a>0}$) \cite{Sachdev99, Carr11}.
Indeed, in panels (a) and (c) of Fig.\,\ref{fig:N=10_energies}, which correspond to ${\mbox{QPT-1}_{+}}$ and ${\mbox{QPT-1}_{0}}$ of Eq.\,\eqref{eq:qptsep}, we see a~sharply closing gap, while in panel (b), which corresponds to ${\mbox{QPT-2}}$, the gap is larger.
In the latter case, we observe the formation of nearly degenerate parity doublets behind the second-order critical point, in accord with the QPT mechanism based on the spontaneous symmetry breaking for ${\chi=0}$.

\begin{figure}[t]
    \centering
    \includegraphics{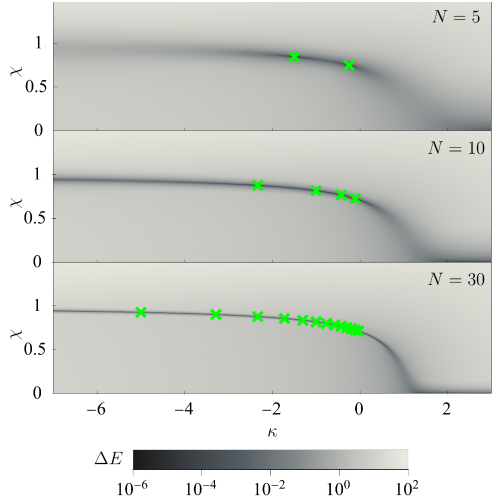}
    \caption{The DPs (red crosses) of Hamiltonian \eqref{eq:hamibst} satisfying Eq.\,\eqref{eq:diabolicPointCoordinates} for ${N=5}$, 10 and 30 (in the latter two cases, a~part of DPs falls to the $\chi<-7$ region, out of the displayed range). The background depicts the $(\kappa,\chi)$-dependence of the energy gap ${\Delta E^{(N)}(\llambda)}$ for the corresponding $N$.}
    \label{fig:singularities}    
\end{figure}

At some isolated points (the DPs) in the parameter plane, the gap $\Delta E^{(N)}(\llambda)$ may exactly vanish even for finite values of $N$.
This actually happens in our model \eqref{eq:hamibst}.
Here, the DPs come in pairs symmetric under the ${\chi\to-\chi}$ reflection. 
The first pair appears for $N=2$, and then the number of pairs $n$ increases according to ${n=\lfloor N/2\rfloor}$, where $\lfloor x\rfloor$ is the floor function.
The DP locations read as follows, 
\begin{equation}
    \bigl(\kappa_{\rm d}^{(l)},\chi_{\rm d}^{(l)}\bigr)=
    \left(\frac{1\!-\!2l}{N\!-\!2l\!+\!1},\pm\sqrt{\frac{N}{2N\!-\!2l\!+\!1}}\right) 
    \label{eq:diabolicPointCoordinates},
\end{equation}
where ${l=1,\dots,n}$ enumerates the pairs in order from the largest (${l=1}$) to the smallest (${l=n}$) value of $\kappa_{\rm d}^{(l)}$.
These expressions were deduced from a~numerical analysis performed for ${N\leq 10}$ (the cases ${N=2,3}$ can be done analytically) and then verified by extensive numerical calculations up to ${N=1000}$. 
The patterns of DPs for three particular values of $N$ are shown in Fig.\,\ref{fig:singularities}.

The formula \eqref{eq:diabolicPointCoordinates} implies that the first and the last DPs (the ones most on the right and most on the left in~$\kappa$) are located at 
\begin{equation}
\begin{array}{rcl}
\bigl(\kappa_{\rm d}^{(1)},\chi_{\rm d}^{(1)}\bigr)
\!\!\!&=&\!\!
\quad\bigl(-\frac{1}{N-1},\pm\sqrt{\frac{N}{2N-1}}\,\bigr),
\\
\bigl(\kappa_{\rm d}^{(n)},\chi_{\rm d}^{(n)}\bigr)
\!\!\!&=&\!\!\!
\left\{\begin{array}{ll}
\bigl(1\!-\!N,\pm\sqrt{\frac{N}{N+1}}\,\bigr) & \text{for}\,N\,\text{even,} \\
\bigl(1-\frac{N}{2},\pm\sqrt{\frac{N}{N+2}}\,\bigr) & \text{for}\,N\,\text{odd,}
\end{array}\right.
\end{array}
\label{eq:dps}
\end{equation}
hence in the limit ${N\to\infty}$ the DPs fill the whole interval ${\kappa\in(-\infty,0)}$.
Moreover, it can be checked that the DP coordinates $(\kappa_{\rm d}^{(l)},\chi_{\rm d}^{(l)})$
for all $l$ and any $N$ satisfy the formula for the $\mbox{QPT-1}_{\pm}$ phase separatrices in Eq.\,\eqref{eq:qptsep}.
One can, therefore, picture the genesis of these $N\to\infty$ separatrices in the ${\kappa<0}$ domain in terms of the two chains of DPs that get more and more condensed as $N$ increases.
However, the ${\kappa>0}$ domain shows no DPs, so the QPT mechanism in this domain must be different.
We note that the only ground-state related DP in this domain is located at the asymptotic point ${(\kappa,\chi)\to(+\infty,0)}$, where the parity doublet becomes degenerate for Hamiltonian~\eqref{eq:hamibst} with any finite $N$.


\begin{figure*}[t]
    \centering
    \includegraphics{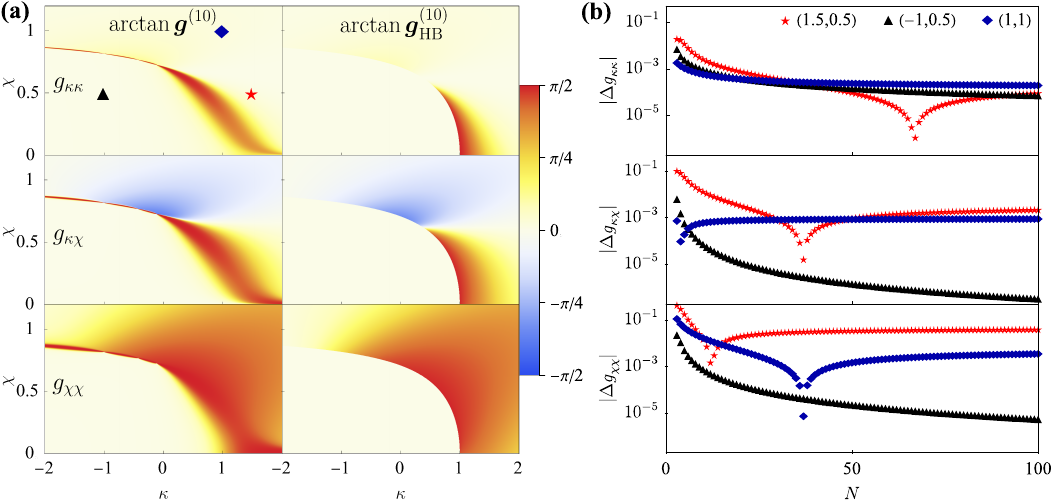}
    \caption{A comparison of the exact and HB metric tensors for the ${f=1}$ boson Hamiltonian \eqref{eq:hamibst}. (a)~Components of the exact metric tensor $\bm{g}^{(N)}(\kappa,\chi)$ and of the HB approximation $\bm{g}_{\rm HB}^{(N)}(\kappa,\chi)$ from Eq.\,\eqref{eq:metr} for ${N=10}$. The $\arctan$ transformation is used to gain higher contrast. (b)~The absolute difference ${\Delta\bm{g}=|\bm{g}^{(N)}\!-\bm{g}_{\rm HB}^{(N)}|/N}$ (the three components) as a~function of~$N$ at three $(\kappa,\chi)$ points specified in the legend and marked by the respective bullets in the top left image. We see that the intensive metric tensor fails to converge to the HB approximation with increasing~$N$.}
    \label{fig:metric_HB_10}
\end{figure*}

\section{Ground-state geometry of the bosonic model}
\label{sec:geomodel}

\subsection{Geometry in the full parameter space}

We will now analyze the geometry of the ${f=1}$ boson model from Sec.\,\ref{sec:LMG}.
At first, we describe global geometric properties of the ground-state manifold within the entire parameter space ${\llambda\equiv(\lambda_1,\lambda_2)\equiv(\kappa,\chi)\subseteq\R^2}$, i.e., across various quantum phases separated by the QPT critical borderlines~\eqref{eq:qptsep}.

Let us start with the HB approximation of the ground-state geometry.
As explained in Appendix~\ref{sec:bosco}, geometry of boson condensates \eqref{eq:hbstates} for ${f=1}$ corresponds to the 2-sphere with radius ${r=\sqrt{N/2}}$, see Eq.\,\eqref{eq:sphere}.
The Ricci curvature ${R=8/N}$ of this manifold is positive but converges to zero with ${N\to\infty}$.
Since in our model the infinite-size condensate parameter $\lim_{N\to\infty}\rho^{(N)}_{\rm min}\equiv\rho$ is real, see Sec.\,\ref{sec:phase}, the resulting ground-state geometry is only a~subset of this sphere.
In particular, the HB ground-state manifold in phase~I corresponds to the north pole ${\theta=0}$ of the sphere (the condensate of $s$-bosons with ${\rho=0}$) and the phases $\mbox{II}_+$ and $\mbox{II}_-$, respectively, are represented by the ${\phi=0}$ and ${\phi=\pi}$ meridians with $\theta\in[0,\pi)$.
In accord with Fig.\,\ref{fig:geophases}, transitions between any pair of phases can be realized in a~discontinuous way via one of the first-order QPTs, but the existence of the point ${\theta=0}$ connecting all three geometries guarantees that such a~transition can also be performed in a~continuous way, via the second-order QPT. 

It is obvious that any non-redundant parametrization of this trivial ground-state geometry has dimension ${D=1}$, while all ${D>1}$ parametrizations must be degenerate.
Indeed, the HB approximation of the metric tensor, see Eq.\,\eqref{eq:metrcondlam}, reads as
\begin{equation}
        \bm{g}_{{\rm HB}}^{(N)}(\kappa,\chi)=\frac{N}{(1+\rho^2)^2}        
        \begin{pmatrix}
            (\partial_\kappa\rho)^2 & (\partial_\kappa\rho)(\partial_\chi\rho)\\
            (\partial_\chi\rho)(\partial_\kappa\rho) & (\partial_\chi\rho)^2
        \end{pmatrix},
\label{eq:metr}
\end{equation}
where the matrix is arranged as ${\bm{g}=\left(\begin{smallmatrix}g_{\kappa\kappa} & g_{\kappa\chi}\\ g_{\chi\kappa} & g_{\chi\chi}\end{smallmatrix}\right)}$ and abbreviated derivative symbols ${\partial_\kappa\equiv\frac{\partial}{\partial\kappa}}$, ${\partial_\chi\equiv\frac{\partial}{\partial\chi}}$ are used.
We immediately see that ${{\rm det}\,\bm{g}_{{\rm HB}}^{(N)}=0}$. 
This follows from the fact that at each point, the distance element~${\d\ell}$ unavoidably vanishes in the direction of the contour ${\rho(\kappa,\chi)={\rm const}}$.
Metric tensor \eqref{eq:metr} is not invertible, and Christoffel symbols, geodesics and scalar curvature in Eqs.\,\eqref{eq:geodesic}, \eqref{eq:christ} and \eqref{eq:ricci} cannot be defined.
The length functional~\eqref{eq:length} is degenerate, assigning the same length to an infinite number of paths between fixed initial and final points in the $(\kappa,\chi)$ plane. 
To overcome this deficiency of the HB description while keeping dimension ${D=2}$ of the parameter space, one would need to consider either an ${f=1}$ model with complex $u_{kl}$ and $v_{klmn}$ in Eq.\,\eqref{eq:hamib}, leading to complex $\alpha(\llambda)$, or a~model with ${f>1}$.
However, this is not needed because, as explained in Sec.\,\ref{sec:genbos}, the actual ${N\to\infty}$ ground state does not coincide with the condensate \eqref{eq:hbstates} and the resulting exact ground-state metric tensor (away of eventual DP and QPT singularities) is expected to be nondegenerate. 

\begin{figure*}
    \centering
    \includegraphics{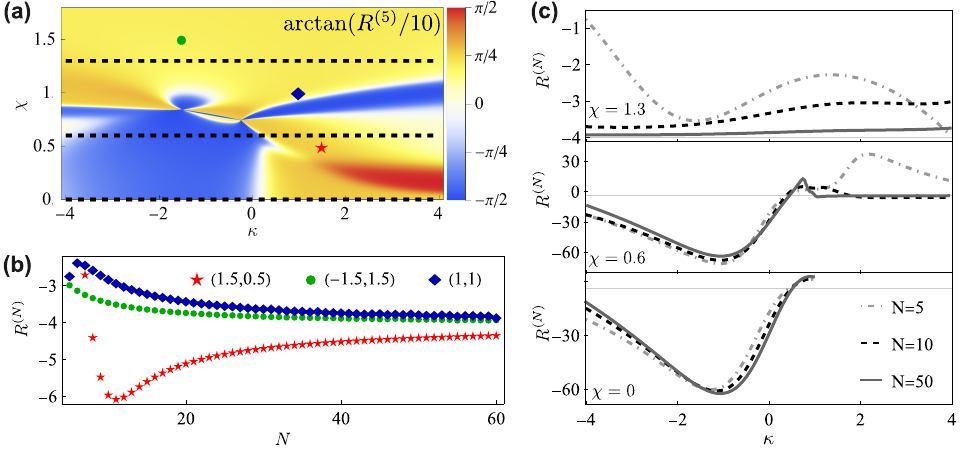}
    \caption{The Ricci scalar extracted from the exact ground-state metric tensor of the ${f=1}$ boson Hamiltonian \eqref{eq:hamibst}. (a)~A~map of $R^{(N)}(\kappa,\chi)$ for ${N=5}$ (the $\arctan$ transformation is used to increase the contrast). (b)~The dependence of $R^{(N)}$ on $N$ at three $(\kappa,\chi)$ points specified in the legend and marked by the respective bullets in panel~(a). In these points, the convergence of~$R^{(N)}$ to $-4$ with increasing $N$ is clearly observed. (c)~Cuts of the $R^{(N)}(\kappa,\chi)$ dependence along three fixed values of~$\chi$ specified in the legend and marked by dashed lines in panel~(a). The cuts are drawn for ${N=5,10}$ and 50, indicating complicated dependence of the Ricci scalar both on the control parameters and on the size.}
    \label{fig:ricci}    
\end{figure*}

The exact metric tensor of Hamiltonian \eqref{eq:hamibst} is compared with its HB approximation in Fig.\,\ref{fig:metric_HB_10}. 
The exact metric tensor (up to some numerical precision) is determined from Eqs.\,\eqref{eq:gRe} and \eqref{eq:geopert} with the actual ground state obtained by numerical diagonalization of the Hamiltonian in the ${j=N/2}$ Hilbert space.
The HB calculation follows Eq.\,\eqref{eq:metr} with $\rho(\kappa,\chi)$ implicitly defined in Eq.\,\eqref{eq:fun}, cf.\,Fig.\,\ref{fig:transitionCompare}(a).
Color plots in Fig.\,\ref{fig:metric_HB_10}(a) compare separately the three independent components of the exact and HB-approximated metric tensors~$\bm{g}^{(N)}$ and~$\bm{g}_{{\rm HB}}^{(N)}$ for ${N=10}$ in a~part of the $\kappa\times\chi$ plane including the ground-state phases~I and~$\mbox{II}_+$ (the domain ${\chi<0}$ would yield mirror-symmetric and -antisymmetric results for diagonal and off-diagonal matrix elements, respectively).
In the maps of the~$\bm{g}^{(10)}$ components we may notice three DPs that appear on the\,$\mbox{ QPT-1}_{+}$ precursor curve.
The graphs in Fig.\,\ref{fig:metric_HB_10}(b) depict the absolute difference $\Delta g_{\mu\nu}$ between intensive metric tensors $g^{(N)}_{\mu\nu}/N$ and $g^{(N)}_{{\rm HB}\,\mu\nu}/N$ at three selected $(\kappa,\chi)$ points (indicated in the upper leftmost panel) as a~function of ${N\leq 100}$.

We observe that the color maps of the metric tensor components exhibit some similarity of the exact and HB results, but differences remain apparent even for large values of~$N$.
As seen in the $\Delta g_{\mu\nu}$ graphs, the HB prediction of the intensive metric tensor does not converge to the exact value for increasing~$N$.
This illustrates the limitations of the HB description of the ground-state wave function outlined in Sec.\,\ref{sec:genbos}.

As anticipated above, the exact finite-$N$ metric tensor in Fig.\,\ref{fig:metric_HB_10}---of course in the parameter domains where it is defined, i.e., away from the DPs---is nondegenerate and therefore allows one to extract the underlying invariant geometry.
The Ricci scalar $R^{(N)}$ describing the invariant curvature of the ground-state manifold derived from the exact metric tensor $\bm{g}^{(N)}$, see Eq.\,\eqref{eq:ricci}, is depicted in Fig.\,\ref{fig:ricci}.
Panel~(a) shows the parameter dependence of~$R^{(N)}(\kappa,\chi)$ for ${N=5}$.
In this case, only two DP singularities are present on the $\mbox{ QPT-1}_{+}$ precursor curve.
Panel~(b) probes the dependence of $R^{(N)}$ on $N$ at three $(\kappa,\chi)$ points indicated in panel~(a). 
Finally, panel~(c) shows the dependence of~$R^{(N)}(\kappa,\chi)$ on $\kappa$ along three ${\chi={\rm const}}$ cuts for three values of $N$.

Figure~\ref{fig:ricci} manifests a~rather complex dependence of the Ricci scalar on both control parameters and on the number of bosons.
A comparison with Fig.\,\ref{fig:metric_HB_10} shows that $R^{(N)}(\kappa,\chi)$ is large and exhibits strong variability with $(\kappa,\chi)$ and~$N$ even in the parameter domains where components of the metric tensor are rather small.
Indeed, following the discussion at the end of Sec.\,\ref{sec:geo}, we point out that since the exact metric tensor $\bm{g}^{(N)}(\kappa,\chi)$ is not far from the degenerate form $\bm{g}_{{\rm HB}}^{(N)}(\kappa,\chi)$, the geometry is very sensitive even to very small variations of the metric-tensor components.
This is further illustrated in Fig.\,\ref{fig:christoffel} showing exact components of the Christoffel symbols, Eq.\,\eqref{eq:christ}, for ${N=10}$.
We see that Christoffel symbols, which underlie all geometrical quantities, including the Ricci scalar, exhibit rather complicated parameter dependence even in the domains far from the QPT borderlines.
Let us note that $\Gamma^{(N)}_{\alpha\beta\gamma}(\kappa,\chi)$ are even or odd functions of parameter $\chi$ for odd or even values of the sum ${\alpha+\beta+\gamma}$, respectively, in accord with the symmetries of our Hamiltonian~\eqref{eq:hamibst}.

What can be deduced from the Ricci scalar?
The sign of $R^{(N)}(\kappa,\chi)$ describes the local flow of geodesic curves near the point $(\kappa,\chi)$ averaged over both mean directions in the parameter plane \cite{Nakahara03}.
The positive or negative signs correspond to predominantly convergent or divergent geodesic flows, respectively, and the zero value represents a~parallel flow associated with flat geometry.
In Fig.\,\ref{fig:ricci}, we observe both signs of the Ricci scalar.
It turns out that everywhere in phase~{II} the function $R^{(N)}(\kappa,\chi)$ converges with increasing~$N$ to constant negative value ${R=-4}$. 
In phase~I the large-$N$ limit of $R^{(N)}(\kappa,\chi)$ is a~$\chi$-independent function $R(\kappa)$ which takes both positive and negative values: With decreasing $\kappa$ this function drops from ${R\approx +2.5}$ at ${\kappa=+1}$ (the QPT-2 point) to ${R\approx -62}$ at ${\kappa=-1}$ and rises for ${\kappa<-1}$, reaching again positive values in the domain of ${\kappa\lesssim-4.7}$. 

\begin{figure}[t!]
    \centering
    \includegraphics{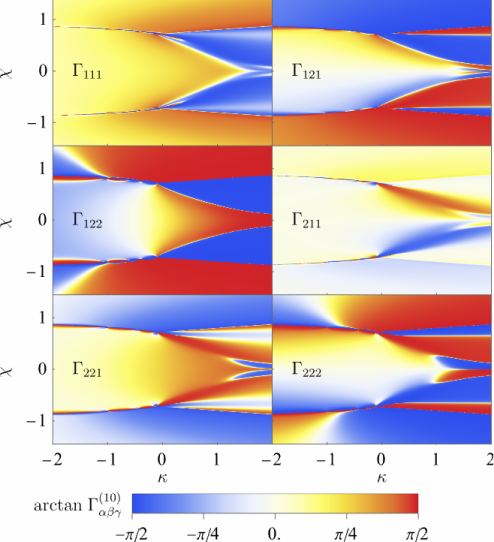}
    \caption{Components of the Christoffel symbols $\Gamma^{(N)}_{\alpha\beta\gamma}$ for Hamiltonian \eqref{eq:hamibst} with $N=10$. The missing two components follow from the relation ${\Gamma^{(N)}_{\alpha\beta\gamma}=\Gamma^{(N)}_{\alpha\gamma\beta}}$.}
    \label{fig:christoffel}    
\end{figure}

Constant negative value of the Ricci scalar within phase~{II} may seem to indicate a~simple invariant geometry associated with that phase in the ${N\to\infty}$ limit, e.g., the geometry associated with the surface of the hyperbolic paraboloid (some bounded 2D objects in 3D space with constant negative curvature, such as the pseudosphere, can be excluded for their cyclic character).
However, it is not possible to make such a~statement just on the basis of the Ricci scalar, without careful analysis of the metric tensor in the whole parameter domain of phase~{II}.
Such an analysis was not performed here.
In any case, phase~I with alternating domains of positive and negative curvature does not allow us even to dream of any elementary geometrical object underlying this phase.
The geometry behind the ground-state manifold in phase~I is apparently rather complex and may not even be represented by any conceivable object embedded in 3D space.
Nevertheless, we note that a~simple geometric representation of the LMG quantum phases may be found in a~higher-dimensional space via an appropriate generalization of the $st$-boson Hamiltonian. 

In this context, let us also point out that our present observations may seem to be in conflict with those of~Ref.\,\cite{Gutierrez21}, where an analogous geometric analysis was performed for another Hamiltonian of the LMG type.
For that Hamiltonian, the ${N\to\infty}$ limit applied in the ground-state phase analogous to our phase~I yields ${R=-4}$, while in the phase analogous to our phase~II the limit was ${R=0}$.
To explain this discrepancy, we must stress that our Hamiltonian~\eqref{eq:hamibst} and the Hamiltonian of Ref.\,\cite{Gutierrez21} represent different subsets of the general LMG parameter space and therefore do not yield the same invariant geometry.
We anticipate that results on the ground-state geometry assigned to various special forms of the LMG model can be unified in a~higher-$D$ general parametrization of the LMG Hamiltonian in the form \eqref{eq:hamib} involving all permissible interaction terms up to a~certain order.
However, this is not the purpose of the present work.

\begin{figure}[t!]
    \centering
    \includegraphics{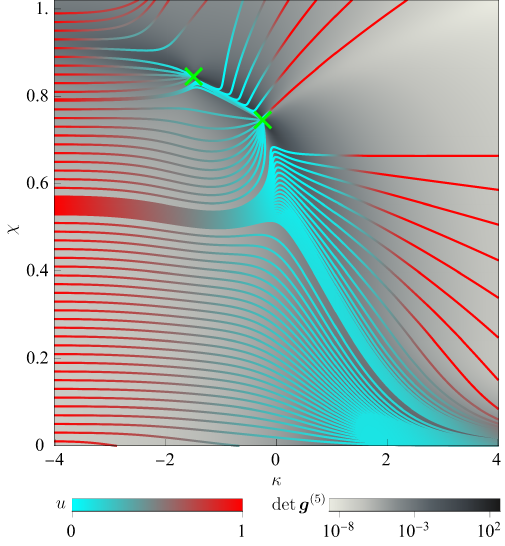}
    \caption{An example of the geodesic flow for the model~\eqref{eq:hamibst} with ${N=5}$. Cauchy boundary condition is applied with initial values ${\llambda=(\kappa,\chi)=(-4,\frac{2k-1}{100})}$, ${k=1,2,\ldots}$ and ${\frac{\d}{\d\tau}\llambda=(1,0)}$. The background shade and curve color encodes the metric tensor determinant and the ordinary speed \eqref{eq:plainspeed}, respectively. Green crosses mark two DPs.} 
    \label{fig:geods}    
\end{figure}

Figure~\ref{fig:geods} shows the flow of geodesic curves in the plane $\kappa\times\chi$ for ${N=5}$. 
In contrast to Fig.\,\ref{fig:geods_illustration}, where we saw geodesics emitted from a~single point in different directions, the present picture draws geodesics emitted from various points (located along the line ${\kappa=-4}$) in the same (horizontal) direction.
The background shade encodes the determinant of the metric tensor, while the color of the geodesic curves represents the ordinary speed~$u$ from Eq.\,\eqref{eq:plainspeed}.
Two DPs are marked by crosses.

We see in Fig.\,\ref{fig:geods} that individual geodesics exhibit very different shapes.
The fate of a~given geodesic depends very sensitively on the initial point from which it is emitted, the essential role being played by absorption and scattering on the DPs.
This is a~consequence of the complex geometry near DPs discussed in Sec.\,\ref{sec:DP}.
While some geodesics in Fig.\,\ref{fig:geods} seem to terminate at the DPs, some others are deviated from the DPs at high angles. 
Some of the geodesics are affected by both DPs. 
The deviated geodesics form a~flow along the region with a~large metric tensor determinant, which appears here as a~precursor of the infinite-size QPT borderline.

One can compare the observed behavior of the geodesics in Fig.\,\ref{fig:geods} with the dependence of the ${N=5}$ Ricci scalar in Fig.\,\ref{fig:ricci}, where the regions with positive or negative values of the Ricci scalar correspond, respectively, to predominantly convergent or divergent flows of geodesic trajectories.
Indeed, a~strongly focusing effect in the geodesic flow is observed in the ${\kappa\gtrsim 1}$ region with ${R^{(N)}(\kappa,\chi)>0}$ near the ${\chi=0}$ axis, while in the regions with ${R^{(N)}(\kappa,\chi)<0}$ inside both phases~I and~{II} the geodesic flow has more or less defocusing (divergent) character.
Near DPs, the focusing and defocusing effects mutually compete, which makes the behavior of geodesics irregular.
In the following, we study the geometrical effects of DPs in more detail.

\subsection{Geometry near DP}
\label{sec:DPboson}

\begin{figure}[t]
    \centering
    \includegraphics{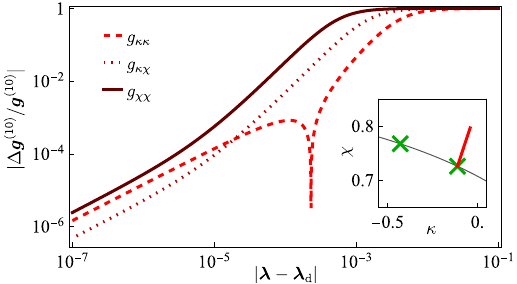}
    \caption{The two-level approximation of the metric tensor near the DP with ${l=1}$, see Eq.\,\eqref{eq:diabolicPointCoordinates}, for Hamiltonian \eqref{eq:hamibst} with ${N=10}$. The error ${\Delta \bm{g}^{(N)}=\boldsymbol{\mathsf{g}}^{(N)}_{{\rm d}}-\bm{g}^{(N)}}$ of the two-level approximation of individual metric tensor components is expressed relative to the exact value for a~path terminating at the DP with angle ${\mathit{\Theta}=\frac{\pi}{4}}$ (red line in the inset). We see that the relative error decreases as the distance $\mathcal{R}$ from the DP tends to zero.} 
    \label{fig:metric_error}    
\end{figure}

\begin{figure*}[t]
    \centering
    \includegraphics[width=\textwidth]{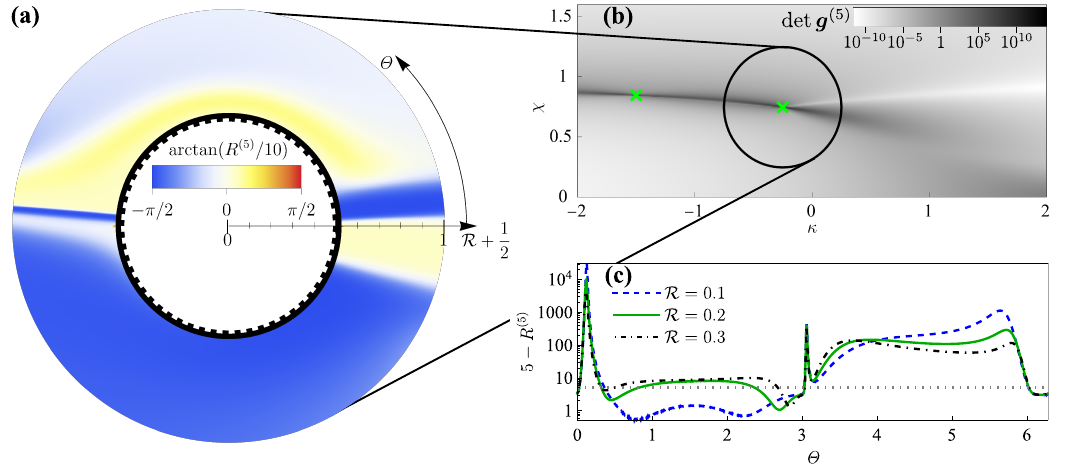}
    \caption{The geometry near the DP with ${l=1}$ for the model \eqref{eq:hamibst} with ${N=5}$.
    (a)~The Ricci scalar $R^{(N)}$ in shifted coordinates ${(\mathcal{R}+\frac{1}{2},\mathit{\Theta})}$. (b)~The metric tensor determinant in a~wider region of $(\kappa,\chi)$, with crosses locating the DPs and the circle marking the area covered in panel~(a). (c)~The dependence of $R^{(N)}$ on $\Theta$ at three different radii. The picture demonstrates a~nontrivial near-DP geometry, which is untraceable by any simple universal model.}
    \label{fig:detg_near_DP}    
\end{figure*}

\begin{figure}
    \includegraphics{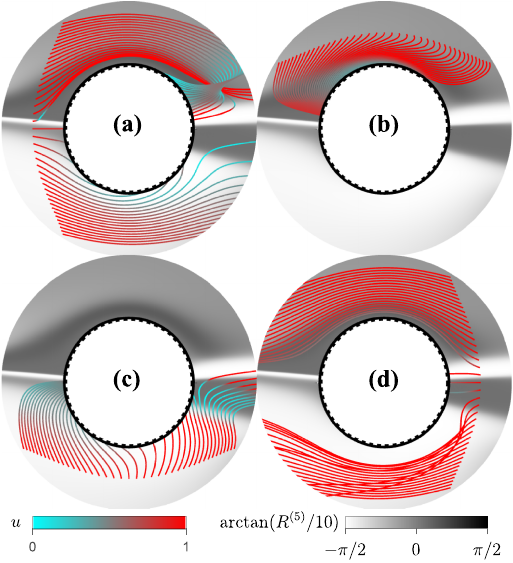}
    \caption{The behavior of geodesics near the DP with ${l=1}$ for the model \eqref{eq:hamibst} with ${N=5}$ (cf.\,Fig.\,\ref{fig:detg_near_DP}). The geodesics are emitted (a) from the vertical line ${\kappa=\kappa_{\rm d}-0.25}$ in the direction to the right, (b) from the horizontal line ${\chi=\chi_{\rm d}+0.25}$ in the downward direction, (c) from the horizontal line ${\chi=\chi_{\rm d}-0.25}$ in the upward direction, and (d) from the vertical line ${\kappa=\kappa_{\rm d}+0.25}$ in the direction to the left. In each panel, we plot 41 geodesics which equidistantly sample the initial line segment of ordinary length ${s=0.8}$, placed symmetrically with respect to the DP. The background shade and color of geodesics encode the Ricci scalar $R^{(N)}$ and the ordinary speed $u$, respectively.}
    \label{fig:geodesics_zoom}
\end{figure}

The ${f=1}$ boson model gives us an excellent opportunity to perform realistic case studies of the behavior of various geometrical quantities near a~DP.
We first focus on the two-level approximation of the metric tensor presented in Sec.\,\ref{sec:DP}, see Eq.\,\eqref{eq:xyz}. 
Just for the sake of completeness, we note that the reality of matrix elements of our Hamiltonian \eqref{eq:hamibst} implies that parameter~$y$ of the two-level Hamiltonian \eqref{eq:ham2} can be constantly set to zero.

Figure~\ref{fig:metric_error} shows the relative errors of individual components of the near-DP metric tensor~$\boldsymbol{\mathsf{g}}^{(N)}_{{\rm d}}(\kappa,\chi)$ determined by the two-level approximation with respect to the exact value $\bm{g}^{(N)}(\kappa,\chi)$ following from the numerical diagonalization of the $N$-boson model.
The calculation was done for a~particular straight path terminated at the first DP (the rightmost one) in the $\kappa\times\chi$ plane for the system with ${N=10}$.
We notice a~clear overall convergence (polynomial close the the DP) of the two-level approximation to exact components as the ordinary distance ${\mathcal{R}=|\llambda-\llambda_\mathrm d|}$ from the DP decreases.
This indicates the validity of the two-level approximation near the DP as far as the components of the metric tensor are considered.
We note that the relative error for one of the components, namely $\Delta g^{(N)}_{\kappa\kappa}/g^{(N)}_{\kappa\kappa}$, shows a~dip at a~distance where the straight path accidentally crosses the zone with ${\Delta g_{\kappa\kappa}(\kappa,\chi)=0}$ (paths approaching the DP from other directions may cross similar zones for the other metric tensor components).

Despite the validity of the two-level method for approximations of the metric tensor close to the DP, the previously gained insight suggests that the prediction of the near-DP geometry is a~much more difficult discipline.
This statement is clearly documented in Fig.~\ref{fig:detg_near_DP}, which shows the Ricci scalar $R^{(N)}(\kappa,\chi)$ near the DP in shifted polar coordinates $(\mathcal{R}+\mathcal{R}_0,\mathit{\Theta})$ with ${\mathcal{R}_0=\frac{1}{2}}$.
The calculation is now done for ${N=5}$.
The map of the parameter dependence of the Ricci scalar is presented in panel~(a), and three ${\mathcal{R}=\mathrm{const}}$ cuts through this dependence are shown in panel~(c).
We stress that in panel (a), the DP itself is represented by the inner circular boundary, which correctly visualizes the dependence of the geometry in the ${\mathcal{R}\to 0}$ region on angle $\mathit{\Theta}$.
The region inside this boundary has no meaning.
Panel~(b) shows a~zoomed-out view of the metric tensor determinant in the $\kappa\times\chi$ plane, where the gently sloping dark blur, associated with the region with large determinant values, represents a~finite-size precursor of the first-order QPT.

Figure~\ref{fig:detg_near_DP} discloses a~rather complex dependence of the Ricci scalar on the coordinates $\mathcal{R}$ and $\mathit{\Theta}$. 
We observe alternating regions with positive and negative values of~$R^{(N)}$, including low ridges and gradual valleys as well as two very deep gorges of negative values (sharp positive peaks in the plot of ${5-R^{(N)}}$).
Since the metric tensor very close to the DP becomes nearly degenerate, as we know from Sec.\,\ref{sec:DP}, the geometry on the inner circle with ${\mathcal{R}\approx 0}$ in Fig.~\ref{fig:detg_near_DP}(a) remains in fact undetermined.
We can say that at the inner circle the geometry collapses.
Although the invariant geometric quantities at any small distance above this circle are in principle defined, from the discussion at the end of Sec.\,\ref{sec:geo}, we know that they are unstable and most likely wildly varying, cf.\,Fig.~\ref{fig:detg_near_DP}(c).
This instability strongly affects the evolution of geodesics which enter the near-DP domain.

Figure~\ref{fig:geodesics_zoom} depicts the behavior of geodesics near the DP from Fig.\,\ref{fig:detg_near_DP}.
The geodesics in individual panels approach the DP from four mutually perpendicular directions.
In particular, they start with the initial coordinate $\llambda|_{\tau=0}$ and initial speed $\frac{d}{d\tau}\llambda|_{\tau=0}$ in the ${\llambda=(\kappa,\chi)}$ parameter space taking the following values: (a) initial coordinates located along a~vertical line segment on the left of the DP, initial speed equal to~$(1,0)$, (b) initial coordinates located along a~horizontal line segment above the DP, initial speed equal to~$(0,-1)$, (c) initial coordinates located along a~horizontal line segment below the DP, initial speed equal to~$(0,1)$, and (d) initial coordinates located along a~vertical line segment on the right of the DP, initial speed equal to~$(-1,0)$.

An inspection of Fig.\,\ref{fig:geodesics_zoom} confirms that near the DP, one finds geodesics with two distinct fates.
A large part of geodesics terminates at the DP.
In some cases, quite surprisingly, the geodesic first bypasses the DP and only then gets to a~descending trajectory.
In contrast, some other geodesics are deflected away from the DP.
This applies for a~fraction of geodesics in panels (a)--(c) and for a~great majority of geodesics in panel (d).
The escape from the DP can happen at earlier or later stages of the geodesic evolution and at different angles.
The observed variety of scenarios and the sensitivity to initial conditions make the geodesic trajectories near the DP analogous to trajectories of a~classically chaotic system.
Note that chaos in the flow of geodesics was first anticipated in Ref.\,\cite{Berry88}.
For related studies of geometry-induced chaos in the framework of classical and relativistic mechanics, see, e.g., Refs.\,\cite{Bala86,Seme10}.

Since, as explained above, the geometry very close to the DP is unstable, numerical issues in the determination of near-DP geodesics have to be carefully addressed.
Indeed, it is clear that the geodesic equation \eqref{eq:geodesic}, which contains Christoffel symbols depending on the metric tensor inverse, cannot even be properly formed in the domain infinitesimally close to the DP.
Nevertheless, the fact that in this domain, the matrix element of the metric tensor in the radial direction drops to zero defines geodesics in this direction even without the need to solve the geodesic equation.
What makes the near-DP calculations of geodesics with arbitrary initial conditions really difficult is that simultaneously with the vanishing of the radial metric tensor component, the component corresponding to the tangential direction diverges.
In spite of these problems, we are confident that all the geodesics shown in Fig.\,\ref{fig:geodesics_zoom} are determined with sufficient accuracy. In our case, the numerical instabilities arise in areas with $\det \mathbf{g}^{(N)}\lesssim 10^{-10}$ and can be easily identified by the noisy behavior of geometric properties. This happens in areas extremely close to the DP, which we exclude from our plot range.

We know that the key feature of the first-order QPT is that no geodesic can pass through it \cite{Kumar12}.
Considerations related to Figs.\,\ref{fig:geods} and \ref{fig:geodesics_zoom} lead us to propose two alternative mechanisms of how such a~first-order QPT boundary in the parameter space arises when the system size grows to infinity.
The first mechanism relies on the gradual formation of a~barrier composed of DPs.
The DPs accumulate in a~chain, or more generally, within a~${(D\!-\!1)}$-dimensional domain, following the region with reduced energy gap $\Delta E^{(N)}$ and the density of DPs in this domain grows with increasing $N$.
This is exactly the case of the $\mbox{QPT-1}_{\pm}$ critical borderlines of our model in the interval ${\kappa\in(-\infty,0)}$, see Eq.\,\eqref{eq:dps} and Fig.\,\ref{fig:singularities}.
In Fig.\,\ref{fig:geods}, we saw a~very early stage of development of this segment of the borderline.
We may anticipate that in the ${N\to\infty}$ limit, the dense formation of DPs becomes totally impassable for geodesics and, therefore, constitutes a~real first-order QPT phase separatrix.

The second mechanism of creating a~non-penetrable boundary in the parameter space is to make all geodesics deflect from or slide along this boundary.
In our bosonic model, this seems to be the case of the ${0\lesssim\kappa\leq 1}$ segment of the $\mbox{QPT-1}_{\pm}$ borderlines and of the whole $\mbox{QPT-1}_{0}$ borderline with ${\kappa\in(1,\infty)}$. 
Indeed, no DPs are located in this domain, and all geodesics slide along the respective critical curves.
For ${\kappa\to 1}$ the $\mbox{QPT-1}_{\pm}$ borderlines merge in the second-order critical point $\mbox{QPT-2}$, see Eq.\,\eqref{eq:qptsep}, to which a~large part of the geodesic flow from phase~I is focused.
We know that discontinuities at second-order QPT can be removed by the transformation \eqref{eq:smoothen}.
Hence the $\mbox{QPT-2}$ point represents the only place where the geodesics can pass from phase~I to phases $\mbox{II}_+$ or $\mbox{II}_-$ in the ${N\to\infty}$ limit.

Although the above conclusions are specific to our present model, similar phase diagrams (with a~second-order QPT located at the intersection of several first-order QPTs) are known also for other many-body systems; see, e.g., Ref.\,\cite{Cejnar07}.
It would be interesting to calculate the distribution of DPs along the QPT separatrices in these systems and learn whether similar geometrical considerations as those presented above can be applied.


\section{Conclusions}
\label{sec:conc}

We studied the geometry of the ground-state manifold associated with interacting $N$-body quantum systems depending on some control parameters $\llambda$.
Our focus was set on the geometrical consequences of QPTs, i.e., nonanalytic changes of the ground state energy and wave function with~$\llambda$, that may emerge in the ${N\to\infty}$ limit of such systems.
We analyzed critical features of the infinite-size geometry, as well as the ways in which these limiting features are being gradually established in finite systems.
We described a~simple system composed of two types of scalar bosons interacting through a~two-body Hamiltonian, in which the general statements about the finite-size precursors of the QPT geometry were exemplified and tested.

It was argued that the infinite-size geometry can be defined only in terms of the intensive metric tensor, in which the decrease of near-eigenstate overlaps induced by an increasing number of system constituents is compensated by appropriate scaling.
It is clear that in the ${N\to\infty}$ limit, the first-order QPT boundary, where the ground state changes with $\llambda$ in a~discontinuous way, constitutes an irremovable geometric singularity impenetrable for all geodesics.
On the other hand, the boundary of a~second-order, or more generally continuous QPT represents a~much softer type of singularity, whose geometrical effects can be (at least partially) eliminated by a~suitable transformation of parameters~$\llambda$. 

General finite-size precursors of criticality in many-body systems were identified with (i)~the closing energy gap ${\Delta E^{(N)}\to 0}$ between the ground and the first-excited states (with increasing~$N$, the gap is known to follow an exponential or polynomial decrease for the first-order or continuous QPT, respectively) and (ii)~exact degeneracies ${\Delta E^{(N)}=0}$ at some isolated DPs in the parameter space (these points may appear along a~finite-$N$ indication of the QPT boundary).
Both these precursors were shown to have a~strong impact on the finite-size geometry. 

The closing energy gap, in general, leads to an increase in the metric tensor determinant.
The resulting prolongation of geometric distances then causes deflection of the geodesic flow from the critical boundary.
Hence the domains with reduced values of ${\Delta E^{(N)}}$ precede the formation of the geometric barriers associated with infinite-size QPTs while keeping well-defined geometry for all finite values of~$N$. 

On the other hand, isolated DPs with ${\Delta E^{(N)}=0}$ represent true finite-$N$ singularities, whose role in quantum geometry can be compared to the role of black holes in the geometry of spacetime.
While the geometry right at the DP \lq\lq collapses\rq\rq\ (is not defined), in a~near vicinity of the DP, one finds practically unpredictable variations of all geometrical quantities.
This again leads to rather intricate behavior of the geodesics.
In particular, geodesics entering the near-DP parameter domain can be scattered by large angles, which sensitively depend on the initial geodesic direction. 
Thus the DPs can be seen as \lq\lq seeds\rq\rq\ generating chaotic flows of geodesics.
Some other geodesics in the near-DP domain happen, at some stage of their evolution, to follow descending trajectories that terminate at the singularity.
 
Based on these observations, we hinted at two mechanisms that can lead to the emergence of the singular geometry associated with the first-order QPT.
The first mechanism relies on a~gradual formation of a~barrier composed of DPs.
We presented an example of such behavior for a~part of the first-order QPT borderline in our simple two-boson model.
The condensation of DPs along the QPT borderline makes it effectively impassable for geodesics. 
It can be anticipated that in the ${N\to\infty}$ limit, all geodesics approaching sufficiently close to this part of the QPT borderline are finally absorbed by some of the densely chained DPs.

The second mechanism applies along the DP-free part of the first-order QPT borderline of our bosonic model.
The geodesics approaching this part of the borderline are deviated so that they never cross it.
In our bosonic model, this is partly realized with the aid of the point of a~second-order QPT in the intersection of three first-order phase separatrices.
The positive scalar curvature near this point focuses the geodesics so that they all pass to the other side of the phase separatrix through the narrow neck of the second-order QPT.

Last but not least, we demonstrated that although the quantum metric tensor on the ground-state manifold of a~many-body system may be relatively easy to calculate or estimate, the geometry derived from this tensor is a~much more fragile object.
First, the quantities that determine invariant geometry (here, the Christoffel symbols and the Ricci scalar) depend on the first and second derivatives of the metric tensor components, which are sensitive to deviations from the exact behavior.
Second, the inversion of the metric tensor, which also enters all relevant expressions, is unstable if the metric tensor is close to degeneracy.
Thus small perturbations of any nearly-degenerate metric tensor lead to huge variations of geometry, which render any estimates of geometry practically impossible.
This important feature was encountered here in the two-level approximation of the near-DP geometry, as well as in the HB approximation of the global ground-state geometry in terms of boson condensates. 

The present work was intended as a~detailed case study of geometry in many-body systems with precursors of QPTs.
The model-independent part of our analysis resulted in some general conclusions, but maybe the most interesting part of this work was connected to the specific model example.
The observations and surmises formulated on the basis of our model need to be tested in other systems and further supported by general arguments.
From our perspective, the most appealing results of the present work are those on the near-DP quantum geometry.
We plan to address this problem in a~more detailed and comprehensive way in the near future.

\begin{acknowledgments}
We thank Jakub Novotný for valuable discussions. We acknowledge the financial support of the Grant Agency of Charles University under grant no. GAUK 207723.
\end{acknowledgments} 

\appendix

\section{Geometry of bosonic condensates}
\label{sec:bosco}

Here, we analyse the geometry associated with bosonic condensate states \eqref{eq:hbstates} and deduce the HB approximation of the metric tensor on the ground state manifold of a~general bosonic system of the type \eqref{eq:hamib}.
Indeed, in the domains away from eventual ground-state degeneracies, the set of condensate parameters $\aalpha_{\rm min}^{(N)}(\llambda)$ obtained via minimization of the intensive energy functional \eqref{eq:Emin} can be considered as a~natural parametrization of the ground state manifold of the interacting boson system for any~$N$.
In the ${N\to\infty}$ limit, this provides a~unique mapping 
\begin{equation}
    \llambda\mapsto\aalpha(\llambda)\equiv\lim_{N\to\infty}\aalpha_{\rm min}^{(N)}(\llambda),
    \label{eq:conpara}
\end{equation}
where $\alpha_k(\llambda)=\rho_k(\llambda)e^{\iu\phi_k(\llambda)}$ with ${k=1,\dots,f}$, see Eq.\,\eqref{eq:condensate}. 
What geometry is assigned to this $2f$-dimensional parametrization? 

The metric tensor associated with the states \eqref{eq:hbstates} for~$N$ bosons can be written as a~real matrix of the block-diagonal form
 \begin{equation}
    g_{\mu\nu}^{(N)}(\aalpha)\equiv \bm{g}^{(N)}(\aalpha)=N\begin{pmatrix}
        \bm{g^{\rho}}(\aalpha) &0\\
        0&{\bm{g^{\phi}}}(\aalpha)
    \end{pmatrix},
    \label{eq:blockdiag}
 \end{equation}   
where the $f$-dimensional submatrices $g^{\bm{\rho}}_{kl}$ and $g^{\bm{\phi}}_{kl}$ for $\bm{\rho}=(\rho_1,\dots,\rho_{f})$ and $\bm{\phi}=(\phi_1,\dots,\phi_{f})$, respectively, read as
\begin{eqnarray}
    g^{\bm{\rho}}_{kl}(\aalpha)&=&\frac{S(\aalpha)\delta_{kl}-\rho_k\rho_l}{S(\aalpha)^2},
    \label{eq:condmetr1}\\
    g^{\bm{\phi}}_{kl}(\aalpha)&=&\frac{S(\aalpha)\rho_k^2\delta_{kl}-\rho_k^2\rho_l^2}{S(\aalpha)^2}.
\label{eq:condmetr2}
\end{eqnarray}
So the squared distance between condensates $\ket{\psi_{\rm HB}^{(N)}(\aalpha)}$ and ${\ket{\psi_{\rm HB}^{(N)}(\aalpha\!+\!\d\aalpha)}}$ is 
\begin{equation}
 \d\ell^2=N\left[g^{\bm{\rho}}_{kl}(\aalpha)\d\rho_k\d\rho_l+g^{\bm{\phi}}_{kl}(\aalpha)\d\phi_k \d\phi_l\right].  
\end{equation}

The Ricci scalar curvature following from the metric \eqref{eq:condmetr1} and \eqref{eq:condmetr2} is always constant and positive, equal to 
\begin{equation}
    \Ric=\frac{4f(f+1)}{N}.
    \label{eq:constcurva}
\end{equation}
We note that the corresponding manifold becomes asymptotically flat as ${N\to\infty}$.
For systems with ${f+1=2}$ types of bosons, the value \eqref{eq:constcurva} implies that the underlying geometry is that of a~2-sphere with radius ${r=\sqrt{N/4}}$.
Indeed, using a~simplified notation ${\rho_1=\rho},{\phi_1=\phi}$, we obtain
\begin{equation}
\d\ell^2=N\frac{\d\rho^2+\rho^2\d\phi^2}{(1+\rho^2)^2}=\frac{N}{4}\bigl(\d\theta^2+\sin^2\theta\,\d\phi^2\bigr),
\label{eq:sphere}
\end{equation}
where in the second equality, we applied transformation
\begin{equation}
\frac{1}{\sqrt{1+\rho^2}}=\cos\frac{\theta}{2}
\end{equation}
from variable ${\rho\in[0,\infty)}$ to angle ${\theta\in[0,\pi]}$. 
Formula~\eqref{eq:sphere} represents the distance on a~2-sphere of the above radius with $\theta$ and $\phi$ associated with ordinary spherical angles.
For ${N=1}$, the formula agrees with the distance on the ground-state manifold of the two-level Hamiltonian \eqref{eq:ham2}, see Eq.\,\eqref{eq:gdia}, which is not an accident since the dimension of the ${f=1}$ model Hilbert space for ${N=1}$ is ${d=2}$.

Except at the QPT critical boundaries, the dependence $\aalpha(\llambda)$ from Eq.\,\eqref{eq:conpara} together with expressions \eqref{eq:blockdiag}--\eqref{eq:condmetr2} allows one to construct the metric tensor on the ground-state manifold in terms of the original parameters $\llambda$.
This HB approximation of the metric tensor reads as
\begin{align}
\bm{g}_{\rm HB}^{(N)}(\llambda)\equiv g_{{\rm HB}\,\mu\nu}^{(N)}=N \biggl[& 
\mbox{Im}\!\!\sum_{k=1}^{f}\alpha^*_k(\partial_{\mu}\alpha_k)\,
\mbox{Im}\!\!\sum_{l=1}^{f}\alpha_l(\partial_{\nu}\alpha^*_l)
\nonumber\\
& -\mbox{Re}\!\!\sum_{k=1}^{f}\alpha^*_k(\partial_{\mu}\partial_{\nu}\alpha_k)
\biggr],
\label{eq:metrcondlam}
\end{align}
where we hide, on the right-hand side, the dependence on~$\llambda$ and use the star for complex conjugation.

An important question is to what extent formula \eqref{eq:metrcondlam} approximates the actual ground-state metric tensor from Eq.\,\eqref{eq:distanc}.
The answer depends on the system and parameter domain of interest, but in general, we can say that the approximation is imperfect in both finite- and infinite-$N$ cases.  
In Sec.\,\ref{sec:genbos}, we pointed out that the boson condensate \eqref{eq:hbstates} constitutes only a~rough representation of the actual ground state.
The overlap of exact ground states at close parameter points is determined not only by the overlap of the corresponding condensates but also by the contributions of $n_{\perp}>0$ states in Eq.\,\eqref{eq:exstates}.
This often results in finite deviations of the exact intensive metric tensor from its mean field estimate, deviations that do not vanish even in the $N\to\infty$ limit.

\end{document}